\documentclass{aa}
\usepackage{amsmath,amssymb,graphics}
\usepackage{times}

\allowdisplaybreaks

\newcommand{\vr}{{v_r}}

\newcommand\vo[1]{{#1}_{\text{H}}}

\newcommand{\zav}[1]{\left(#1\right)}
\newcommand{\hzav}[1]{\left[#1\right]}

\newcommand{\rad}{\mathrm{rad}}

\newcommand{\erf}{\text{erf}}
\newcommand{\Msun}{\ensuremath{\text{M}_{\sun}}}

\newcommand{\de}{\mathrm{d}}
\newcommand{\Teff}{\mbox{$T_\mathrm{eff}$}}

\newcommand{\rs}{\ensuremath{R_\star}}

\newcommand\zm[1]{{#1}}

\begin{document}

\title{The 
winds
of
the
hot massive first stars}
\titlerunning{The
winds
of 
the
hot massive first stars}

\author{J. Krti\v{c}ka\inst{1}
\and J. Kub\'at\inst{2}}
\authorrunning{J. Krti\v{c}ka and J. Kub\'at}

\offprints{J. Krti\v{c}ka, \email{krticka@physics.muni.cz}}

\institute{\'Ustav teoretick\'e fyziky a astrofyziky, 
	P\v r\'\i rodov\v edeck\'a\ fakulta Masarykovy univerzity,
	CZ-611 37 Brno, Czech Republic
	\and
	Astronomick\'y \'ustav, Akademie v\v{e}d \v{C}esk\'e
	republiky, CZ-251 65 Ond\v{r}ejov, Czech Republic}

\date{Received 25th April 2005}

\abstract{We study dynamical aspects of circumstellar environment around
massive zero-metallicity first stars.
For this purpose we apply our NLTE wind models.
We show that the
hydrogen-helium
stellar wind from stationary massive first generation
(Population III) stars (driven either by the line (bound-bound) or
continuum (bound-free and free-free) transitions) is unlikely.
The possibility of expulsion of chemically homogeneous wind and the role
of minor isotopes are also discussed.
Finally, we estimate the importance of hydrogen and helium lines for
shutting off the initial accretion onto first stars and its influence on
initial mass function of first stars.
}

\maketitle

\section{Introduction}

The Big Bang nucleosynthesis has left our universe completely metal-free
(e.g. Coc et al. \cite{coc}).
Thus, at the dawn of the universe there existed a population of stars
composed from hydrogen and helium only.
Numerical simulations show that the physical properties and the
evolution of these first stars (Population III stars) were very
different from the
stellar evolution and properties of Population I and II stars.
Simulations of first star formation (cf. Bromm et al. \cite{bcl},
Nakamura \& Umemura \cite{naum}) show that due to the less efficient
cooling the first stars were probably much more massive than the present
stars. First stars with masses of the order $100\,{\Msun}$
might have been formed.
However, there are indications that the initial mass function of first
stars was bimodal (Nakamura \& Umemura \cite{naum}, Omukai \& Yoshii
\cite{nemukaj}), and thus solar mass first stars may have been formed as
well.

There are no available observations of first stars up to now.
Although the recently discovered extremely low metallicity star
HE~0107--5240 (Christlieb et al. \cite{starka}) was claimed to be the
first star (Shigeyama et al. \cite{sty}), there are some problems with
this explanation.
It is difficult to explain the present surface chemical composition of
this star (especially that of carbon and nitrogen) on the basis of
initial pure hydrogen and helium composition.
For example, Shigeyama et al. (\cite{sty}) explain the CN composition by
the surface enrichment due to the central nucleosynthesis, however there
is a question whether significant amount of CN elements may occur on the
surface for this stellar type according to evolutionary calculations of
Marigo et al. (\cite{mari}).
It is more likely that this star was formed from the material which was
already enriched by the first stars (Bromm \& Larson \cite{brola})
or by subsequent stellar populations.

Consequently, it is necessary to constraint ourselves only to indirect
observation of the first stars.
Probably the most reliable indirect observation of the first stars comes
from the polarisation measurement of cosmic microwave background
radiation.
If the first stars
%
ionized the matter in the universe
significantly,
we should observe its imprint on the cosmic microwave background due
to the light scattering on free electrons.
Recent cosmic microwave background polarisation measurement showed that
the universe was reionized $180 \times
10^6$ years after the Big Bang (Kogut
et al. \cite{kohout}).
Detailed numerical analysis shows that this effect can be attributed to
the generation of massive zero-metallicity stars (Cen \cite{cen},
Wyithe \& Loeb \cite{bililo}, Sokasian et al. \cite{sokyoh}).
Moreover, the level of the sky infrared background radiation can also be
explained by the generation of massive first-generation stars
(Magliocchetti et al. \cite{magli}).

The detailed theoretical study of physical properties of the first
generation stars is a challenging task. Especially appealing is the study of 
its stellar winds, since winds might have a significant impact on the first
stars evolution.
Moreover, the study of outflows of the first stars may be important for
the estimate of the duration of supermassive stars formation.
The possibility of supermassive stars formation depends critically on
the metal abundance in the interstellar medium. 
The presence of a relative small amount of metals might inhibit
subsequent formation of supermassive stars (see Bromm \& Larson
\cite{brola} for a review).
Note, for example, that there are indications that
\zm{the}
initial formation of
supermassive stars might have been inhibited by metals expelled due to
the supernova explosions and that direct black hole formation by
implosion is necessary for the explanation of the duration of massive
star era (Ricotti \& Ostriker \cite{jeremias}).
Thus, the interstellar medium had to remain metal free for a significant
amount of time if the stellar wind did not enrich it with nuclearly
processed material.
This shows how important the study of winds from first generation stars
potentially could be.


The outflows from first stars have not been systematically studied up to
now.
First step in this direction was done by Kudritzki (\cite{kudmet}), who
calculated wind parameters suitable for massive extremely-low
metallicity stars.
As was discussed by Kudritzki and also subsequently supported by
detailed numerical modelling by Krti\v cka et al. (\cite{gla}), some of
these stellar winds may suffer from multicomponent effects (see
Krti\v cka \& Kub\'at (\cite{kkii}) for a detailed modelling of
multicomponent effects in the line-driven stellar winds).
However, to our knowledge
the line-driven stellar winds of zero-metallicity stars have
not been discussed so far.
To fill this gap we present the detailed discussion of pure
\element{H}-\element{He} line-driven winds.

\section{Theoretical limits for line-driven winds}

The necessary condition for launching the line-driven stellar wind is
relatively simple.
The radiative force should exceed the gravity force at some point.
Thus, instead of presenting more detailed numerical models we start with
a discussion of this condition and its application to simple situations.
This helps to better understand more complicated behaviour of numerical
models.

\subsection{Minimum radiative force}

The radiative acceleration due to the absorption or scattering of
radiation is given by (Mihalas \cite{mihalas})
\begin{equation}
\label{nebelvir}
g^{\rad}=\frac{4\pi}{c\rho}\int_0^\infty \chi_\nu H_\nu \, \de \nu,
\end{equation}
where $\rho$ is the mass density,
$\chi_\nu$ is the absorption coefficient,
$H_\nu$ is the radiative flux,
and $\nu$ is the frequency.
%
For the calculation of the radiative force in the optically thin
environment, the radiative flux
$H_c(\nu)$
emerging from the stellar
atmosphere can be directly inserted into Eq.~(\ref{nebelvir}).
However, this is not the case of the stellar wind since the radiative
flux $H_\nu$ is influenced by the absorption and emission of the
radiation in the sorrounding
wind
environment in such case.
Thus, the self-shadowing in the given transition has to be accounted
for, preferably by the solution of the radiative transfer equation.
In the case of rapidly accelerating stellar wind a convenient
approximation of the radiative transfer equation solution
is given by the Sobolev approximation (see Sobolev \cite{sobolevprvni}).


From Eq. (\ref{nebelvir}) follows the limit below which
classical line-driven wind is not possible.
In particular, neglecting continuum absorption, for a given set of
lines the radiative force is maximum,
if lines are not self-shadowed and if lines are optically thin, i.e. if
the optical depths $\tau<1$ in lines
(Gayley \cite{gayley}).
%
Thus, for given occupation numbers the line 
acceleration (i.e. line force per unit of mass)
cannot be larger than
\begin{equation}
\label{silazarmax}
g^\mathrm{rad,\, max}=\frac{4\pi^2e^2}
{\rho m_\mathrm{e}c^2}\zav{1-\mu_c^2} \sum_{\mathrm{lines}}
H_c (\nu_{ij})
g_if_{ij}\zav{\frac{n_i}{g_i}-\frac{n_j}{g_j}},
\end{equation}
where now
$H_c (\nu_{ij})$
is the frequency dependent flux
emerging from the star at the stellar radius \rs,
$\zav{1-\mu_c^2}=\rs^2/r^2$ is a correction due to the dilution of
radiation,
$\nu_{ij}$ and $g_if_{ij}$ are the frequency and the oscillator strength
of a given transition $i\leftrightarrow j$, $n_i$, $n_j$ are the
number densities, 
%
and $g_i$, $g_j$ are the
statistical weights of corresponding levels.

Note that $g^\mathrm{rad,\, max}$ is related to the Gayley's parameter
$\bar Q$ (Gayley \cite{gayley}, Puls et al. \cite{pusle}), namely
\begin{equation}
\label{zmijozeljed}
g^\mathrm{rad,\, max}=g_\mathrm{e}^\mathrm{rad}\bar Q,
\end{equation}
where $g_\mathrm{e}^\mathrm{rad}$ is the radiative acceleration on free
electrons,
\begin{equation}
g_\mathrm{e}^\mathrm{rad}=\frac{\sigma_\mathrm{e} L}{4\pi c r^2},
\end{equation}
where $L$ is the stellar luminosity,
\begin{equation}
\label{mezipoli}
\sigma_\text{e}=\frac{n_\mathrm{e}s_\text{e}}{\rho},
\end{equation}
$s_\mathrm{e}$ is the Thomson scattering cross-section,
and $n_\mathrm{e}$ is the electron density.
The line-driven wind is possible only if the total radiative force is
greater than gravity at some point in the stellar atmosphere or
circumstellar envelope.
Thus, the necessary condition for line-driven wind reads
(neglecting the gas-pressure)
\begin{equation}
\label{zmijozelsil}
g^\mathrm{rad,\, max}+g_\mathrm{e}^\mathrm{rad}=
\zav{\bar Q+1}g_\mathrm{e}^\mathrm{rad}>g,
\end{equation}
where $g={G{M}}/{r^2}$ is the gravity acceleration.
Using Eq.~\eqref{silazarmax} for maximal line 
acceleration
and
${1-\mu_c^2}=\rs^2/r^2$
we can obtain a condition for the lowest limit of line-driven wind as
\begin{equation}
\label{zmijozel}
\Gamma+\frac{4\pi^2e^2}{\rho m_\mathrm{e}c^2 G{M}} R_*^2
   \sum_{\mathrm{lines}} H_c (\nu_{ij})
g_if_{ij}\zav{\frac{n_i}{g_i}-\frac{n_j}{g_j}}>1,
\end{equation}
where $M$ is stellar mass and
\begin{equation}
\label{babice}
\Gamma=\frac{\sigma_e L}{4\pi c G{M}}.
\end{equation}
Using the Gayley's parameter, which for our case reads
\begin{equation}
\bar Q = \frac{\pi e^2}{m_ecs_en_e}
\sum_{\mathrm{lines}}
\frac{H_c (\nu_{ij})}{H_t(R_\ast)}
g_if_{ij}\zav{\frac{n_i}{g_i}-\frac{n_j}{g_j}},
\end{equation}
where $H_t(R_\ast)=L/(4\pi R_\ast)^2$ is the total flux at the stellar
surface $R_\ast$, the condition \eqref{zmijozelsil} may be rewritten as
\begin{equation}
\bar Q > \frac{1}{\Gamma} - 1.
\end{equation}
Let us first assume that the stellar surface flux $H_c (\nu_{ij})$ is
constant for a given transition $i\rightarrow j$ throughout the wind,
i.e. that the frequency dependence of $H_c (\nu_{ij})$ due to the
Doppler shift can be neglected.
In such a case the only quantities in Eq.~\eqref{zmijozel} which
depend on radius are the fractions $n_i/\rho$, $n_j/\rho$.
Thus, for a given star 
Eq.\,\eqref{zmijozel} gives a lower limit for $n_i/\rho$ below which the
line-driven wind is not possible. 
Eq.~\eqref{zmijozel} is a necessary condition for launching
the stellar wind.

Another formulation of the
condition for launching the stellar wind may be obtained from the
requirement that the work done by the radiative force (per unit mass) is
greater than the absolute value of gravitational potential.
It is noteworthy that in the case of wind driven purely by optically
thin lines this condition can be easily obtained by the integration of
Eq.\,\eqref{zmijozelsil} from $r=R_\ast$ to infinity if the wind
ionization and excitation state is constant
(note that in this case fraction $n_i/\rho$ is constant for homogeneous
winds)
\begin{equation}
\label{mrzimor}
\frac{\Gamma G{M}}{R_*}+\frac{4\pi^2e^2}{\rho m_\mathrm{e}c^2} R_*
   \sum_{\mathrm{lines}} H_c (\nu_{ij})
   g_if_{ij}\zav{\frac{n_i}{g_i}-\frac{n_j}{g_j}}>\frac{GM}{R_*}.
\end{equation}
Apparently, this is an identical condition to Eq.~\eqref{zmijozel} and
thus Eq.\,\eqref{zmijozel} is also an energy condition for launching the
stellar wind driven purely by optically thin lines.
However, real line-driven stellar winds are accelerated not only by
optically thin lines but by the mixture of optically thin and optically
thick lines.
Thus, a simple condition \eqref{mrzimor} cannot be used in this case.
Moreover, some of the lines which accelerate the stellar wind may be
optically thick in the photosphere.
In such a case, the stellar flux $H_c$ depends on the wind velocity
(and thus also on the radius) due to the Doppler effect and condition
\eqref{mrzimor} cannot be applied.

\subsection{Application to hydrogen}
\label{vodikprik}

The necessary condition for the existence of line-driven winds
\eqref{zmijozel} is especially simple in the case of the
radiative force only due to the hydrogen lines.
This is only a rough approximation even in the case of stars with zero
metallicity for which also the radiative force due to the helium lines
may be important.
However, condition \eqref{zmijozel} at least provides a guess of
occupation numbers for which the line-driven winds of stars at zero
metallicity may exist.
Moreover, adding other hydrogen-like ions is straightforward.

The oscillator strengths for hydrogen are given by the well known
Kramers formula, which neglecting Gaunt factors reads
(Mihalas \cite{mihalas}, Eq.\,(4.78) therein)
\begin{equation}
\label{oscivo}
g_if_{ij}=\frac{64}{3\pi\sqrt{3}}\zav{\frac{1}{i^2}-\frac{1}{j^2}}^{-3}
\frac{1}{i^3j^3}.
\end{equation}
For the 
resonance lines
it may be assumed
\begin{equation}\label{ninjares}
\frac{n_1}{g_1}-\frac{n_j}{g_j} \approx \frac{n_1}{g_1},
\end{equation}
which in fact means neglecting the stimulated emission in all
resonance transitions. Moreover, the contribution of other than resonance lines
to the radiative force can be in many cases neglected.

Thus, in the case of hydrogen, we can obtain from Eq.\,\eqref{zmijozel}
relatively simple
condition for the existence of line-driven wind in the form of
\begin{equation}
\label{zmijozelh}
\Gamma+\frac{128 \pi e^2}{3\sqrt{3}\rho m_\mathrm{e}c^2 G{M}}R_*^2
   \sum_{j} H_c (\nu_{1j}) \zav{1-\frac{1}{j^2}}^{-3}
   \frac{n_1}{j^3}>1. 
\end{equation}
The main contribution in the sum in condition \eqref{zmijozelh} is
due to the resonance lines and the lowest relative occupation number of
the hydrogen ground level $N_{\text{\ion{H}{i}},1}$, for which the
hydrogen-driven wind is possible, is given by
\begin{multline}
\label{nejmennh}
N_{\text{\ion{H}{i}},1}^{\min}=
\hzav{\sum_{j}H_c(\nu_{1j})\zav{j-\frac{1}{j}}^{-3}}^{-1}
\times \\
\frac{3\sqrt{3} m_\mathrm{e}m_\mathrm{H} c^2 G{M}}{128
\pi e^2 X R_*^2} \zav{1-\Gamma},   
\end{multline}
where $m_\mathrm{H}$ is hydrogen atom mass,
$X=n_\mathrm{H}m_\mathrm{H}/\rho$ is the relative hydrogen abundance,
$n_\mathrm{H}$ is hydrogen number density and
$N_{\text{\ion{H}{i}},1}^{\min}=n_{\text{\ion{H}{i}},1}^{\min}/n_\mathrm{H}$
(where $n_{\text{\ion{H}{i}},1}^{\min}$ is the minimum number density of
ground state neutral hydrogen necessary to launch the stellar wind).
Apparently, for stars close to
the Eddington limit ($\Gamma\rightarrow 1$) the minimum relative number
density $N_{\text{\ion{H}{i}},1}$ for launching the hydrogen line-driven
wind is lower.

\subsection{Application to helium}

\ion{He}{ii} is the most important helium ion for radiative driving
since helium is ionized in the stellar winds of hot stars.
Thus, the radiative force due to helium can be calculated in a very
similar way to that due to \ion{H}{i}. Similarly to
the case of \ion{H}{i}, the resonance lines are the only important
\ion{He}{ii} lines for radiative driving
(assuming that helium is singly ionized ionized).
Thus, we can write
%
for \ion{He}{ii} (similarly as for \ion{H}{i})
\begin{equation}
\frac{n_i}{g_i}-\frac{n_j}{g_j}\approx
\left\{\begin{array}{cc}
      \frac{n_i}{2},&i=1,\\
      0,&i>1.
      \end{array}
\right.
\end{equation}
The condition of existence of wind driven by the absorption in
\ion{He}{ii} lines is (see Eq.~\ref{zmijozelh})
\begin{equation}
\label{zmijozelhe}
\Gamma+\frac{128\pi e^2R_*^2}{3\sqrt{3}\rho m_\mathrm{e}c^2 G{M}}
  n_{\text{\ion{He}{ii}},1}
   \sum_{j}H_c (\nu_{1j}) \zav{1-\frac{1}{j^2}}^{-3}\frac{ 1 }{j^3}>1, 
\end{equation}
and the lowest relative occupation number
$N_{\text{\ion{He}{ii}},1}$ for which the helium-driven wind is possible
is
\begin{multline}
\label{nejmennhe}
N_{\text{\ion{He}{ii}},1}^{\min}= \hzav{\sum_{j}H_c (\nu_{1j})
\zav{j-\frac{1}{j}}^{-3}}^{-1}\times\\*
\frac{3\sqrt{3} m_\mathrm{e}m_\mathrm{He}c^2 G{M}}{128\pi  e^2 Y R_*^2}\zav{1-\Gamma},
\end{multline}
where $m_\mathrm{He}$ is helium atom mass,
$Y=n_\mathrm{He}m_\mathrm{He}/\rho$ is the relative helium abundance,
$n_\mathrm{He}$ is helium number density and
$N_{\text{\ion{He}{ii}},1}^{\min}=n_{\text{\ion{He}{ii}},1}^{\min}/n_\mathrm{He}$.
%
The sums in \eqref{zmijozelhe} and \eqref{nejmennhe} converge due
to the asymptotic behaviour of hydrogen-like oscillator strengths,
$g_if_{ij}\sim{1}/{j^3}$ (see Eq.~\eqref{oscivo}).

In many cases the stellar wind is accelerated by lines which are
optically thick in the photosphere and thus the stellar flux $H_c$
for frequencies close to the photospheric line center
depends strongly on frequency.
Since the radiative flux at a given point is shifted due to the Doppler
effect, the optically thin radiative force depends on velocity (and
consequently also on radius).
Even in such a case the conditions (\ref{nejmennh},\ref{nejmennhe}) can
also be used
\zm{to estimate the minimum conditions for line driven winds}.
However, the actual radiative flux $H_c[\nu(\vr)]$, which depends on
wind velocity (and thus also on radius) should be inserted in these
equations.

\subsection{Stellar wind driven by optically thin lines}
\label{kaptenvit}

We have derived the necessary condition for launching the line-driven
stellar wind.
However, there is another question which should be answered.
Is this condition also the sufficient one?
In other words, are the optically thin lines only able to drive stellar
wind?

\subsubsection{Constant ionization fraction}

Let us first assume that the wind ionization and excitation state does
not depend on wind density and on radius and that the stellar surface
flux $H_c$ does not significantly vary with frequency.
In such a case the optically thin radiative 
acceleration
\eqref{silazarmax}
does not depend on wind density.
Thus, also the wind momentum equation
\begin{equation}
\label{pohybten}
\vr\frac{\de\vr}{\de r}=\frac{2a^2}{r}+\frac{a^2}{\vr}\frac{\de\vr}{\de r}-g+
g^\mathrm{thin} +g_\mathrm{e}^\mathrm{rad},
\end{equation}
(where the optically thin radiative 
acceleration
$g^\mathrm{thin}=g^\mathrm{rad,\, max}$,
$a$ is sound speed, and where we assumed isothermal flow)
does not depend on wind density.
It may seem that the optically thin line-force may launch the stellar
wind with an arbitrary mass-loss rate. However, this is not true,
since in such a case either for higher wind densities some of the lines
would become optically thick or the photon tiring effect (Owocki \&
Gayley \cite{unava}) would impose constrains on the mass-loss rate.

\subsubsection{Variable ionization fraction}
\label{promion}

In the case when the wind ionization and excitation state depends on
wind density, the value of wind mass-loss rate
can
be derived from the hydrodynamic equations.
The optically thin momentum equation \eqref{pohybten} has a critical
point for $\vr=a$.
To obtain a smooth wind solution, the regularity condition
\begin{equation}
\label{havraspar} \frac{2a^2}{r}+g^\mathrm{thin}+g_\mathrm{e}^\mathrm{rad}=g
\end{equation}
must be fulfilled at the critical point.
Since wind ionization and excitation state depends on wind density now,
also the optically thin radiative acceleration depends on wind density.
Thus, the critical condition \eqref{havraspar} depends implicitly on the
wind density and using this equation the mass-loss rate of thin winds
can be calculated.
Note that in the common case when the gas pressure term is negligible,
the condition \eqref{zmijozelsil} will be just fulfilled at the critical point.
This again shows the importance of the condition \eqref{zmijozelsil}.

However, this analysis does not tell us anything about the possibility
to drive the wind material out from the stellar gravitational potential
well.
This condition has to be tested using detailed numerical modelling.
On the other hand, more wind properties can be unveiled using a simple
analysis.
The stellar photosphere deep below the wind region shall be
quasi-static.
Thus, to obtain a smooth flow, the radiative force has to be smaller
than the gravity in this region and the condition \eqref{zmijozelsil}
has not to be fulfilled below the critical point.
The change of wind ionization or excitation state then has to assure
that the equality in \eqref{zmijozelsil} is fulfilled at the critical
point.
Finally, in the outer parts of the wind the inequality in condition
\eqref{zmijozelsil} has to hold, i.e. the radiative force has to be
greater than the gravity in the outer wind parts.

\subsubsection{Frequency dependent stellar flux}

The frequency variable stellar flux gives probably the most attractive
possibility to obtain consistent radiative force to drive an optically
thin wind, i.e. radiative force which is lower than the gravity force
below the critical point and higher outwards.
This is caused by the fact that some lines, which accelerate the
stellar wind, may be optically thick in the photosphere and optically
thin in the stellar wind.
In such a case the stellar flux $H_c$ strongly depends on frequency and,
consequently, due to the Doppler effect, also on the wind velocity.
The stellar wind near the star has low velocity and a given line is
accelerated mainly by the weak flux from the line core.
On the other hand, stellar wind with higher velocities is accelerated
mainly by the stronger flux from the line wings and thus the line force
of the outer wind may be higher than that of inner part of the wind
close to the star.
However, this is not as simple as it looks like since higher flux may
also cause higher excitation and ionization and consequently lower the
radiative force. Clearly, detailed numerical calculations
are necessary to test this possibility.

\section{Numerical calculation of thin winds}
\label{tenkovit}

For the numerical tests we used NLTE wind models described by
Krti\v cka \& Kub\'at (\cite{nlte1}).
The computer program allows to solve
hydrodynamic, simplified radiative transfer, and statistical equilibrium
equations in a radiatively driven stellar wind.
However, for the present purpose 
we slightly changed our code.
Since we do not know in advance whether the stellar wind is possible or
not,
the hydrodynamical variables, i.e.
velocities, temperatures, and densities of all wind components (with an
exception of the electron density) are kept fixed.
This enables us to calculate the model occupation numbers and the
radiative force
regardless the existence of the wind and, consequently,
to test the wind existence.

%
The method itself was slightly improved with respect to the published
one.
We included accelerated lambda iterations (cf. Rybicki \& Hummer
\cite{rybhula}) and Ng acceleration (Ng \cite{ng}).
These improvements will be described in detail in a separate paper
(Krti\v cka \& Kub\'at, in preparation).
Moreover, to obtain a correct surface flux we accounted for the Doppler
effect during the calculation of radiative transfer in lines, i.e. we
shifted the stellar surface flux $H_c$ according to the actual wind
velocity (we accounted for the detailed profiles of photospheric lines,
cf. Babel \cite{babelb}).
Finally, to assure a smooth transitions between stellar atmosphere and
wind, the model atmosphere fluxes are taken at the point where the
density is equal to the boundary wind density.

First, we have performed tests whether the radiative force is large
enough to drive stellar wind in optically thin lines.
Although this gives only the necessary condition for stellar wind and
not the sufficient one (and thus further calculations are necessary to
test whether the stellar wind is possible or not), this approach enables
us to better understand more detailed models.
%

\subsection{Minimum number density for line-driven winds}

We performed two different tests of the possibility of a stellar wind of
zero-metallicity stars driven by optically thin lines.
First, we calculated the ionization structure of a possible extended
stellar atmosphere and compared this ionization structure with minimum
number densities for launching a line-driven wind obtained using
Eqs.\,\eqref{nejmennh} and \eqref{nejmennhe}.

\begin{table}[hbt]
\caption{Stellar and wind parameters used to calculate discussed wind models
taken from Kudritzki (\cite{kudmet}).}
\label{porovioni}
\centering
\begin{tabular}{cccccc}
\hline
\hline
\multicolumn{5}{c}{Stellar parameters} &
%
Model
\\ 
$\log\zav{L/\text{L}_{\sun}}$
&$T_\mathrm{eff}$ & $\log g$ & $R_*$ & $T_\text{wind}$ \\
&\relax [K] & CGS & [R $_{\sun}$]
& [K] \\
\hline
 7.03 & $60\,000$ & 3.95 & 30.38 & $49\,000$ &T60 \\
      & $50\,000$ & 3.63 & 43.76 & $39\,000$ &T50\\
      & $40\,000$ & 3.25 & 68.32 & $29\,000$ &T40\\
 6.91 & $60\,000$ & 3.99 & 26.24 & $49\,000$ \\
      & $50\,000$ & 3.68 & 38.06 & $39\,000$\\
      & $40\,000$ & 3.28 & 59.49 & $29\,000$\\
 6.76 & $60\,000$ & 4.04 & 22.29 & $49\,000$ \\
      & $50\,000$ & 3.73 & 32.10 & $39\,000$\\
      & $40\,000$ & 3.34 & 50.12 & $29\,000$\\
 6.57 & $60\,000$ & 4.11 & 17.78 & $49\,000$ \\
      & $50\,000$ & 3.79 & 25.74 & $39\,000$\\
      & $40\,000$ & 3.41 & 40.18 & $29\,000$\\
 6.42 & $60\,000$ & 4.16 & 15.07 & $49\,000$ \\
      & $50\,000$ & 3.85 & 21.71 & $39\,000$\\
      & $40\,000$ & 3.46 & 33.93 & $29\,000$\\
 6.30 & $60\,000$ & 4.21 & 13.11 & $49\,000$ \\
      & $50\,000$ & 3.89 & 18.88 & $39\,000$\\
      & $40\,000$ & 3.50 & 29.51 & $29\,000$ \\
\hline
\end{tabular}
\end{table}


For our detailed discussion we selected only those stars from the
Kudritzki (\cite{kudmet}) list, which have the largest luminosity, i.e.
those which are close to the Eddington limit.
These stars have the largest radiative force due to the light scattering
on free electrons in their atmospheres and thus also the highest chance
to launch the stellar wind.
Detailed models were calculated also for other stars from the
Kudritzki's list, however for these stars we describe the final results
only.
%
Parameters of all studied model stars are given in
Tab.\,\ref{porovioni}.
%

We compared the minimum number densities necessary for launching the
stellar wind with actual number densities calculated by our NLTE code.
Since we do not know in advance whether the radiative force is strong
enough to drive a wind, the velocity structure of our models is given by
an artificial velocity law
\begin{equation}
v(r)=10^{-3}\sqrt{\frac{5}{3}\frac{kT_{\mathrm{eff}}}{m_\mathrm{H}}}+
4\,10^{7}\,\mathrm{cm\,s}^{-1}\frac{r-R_*}{R_*}.
\end{equation}
This velocity law corresponds to the Taylor expansion of the wind 
velocity around the sonic point.
According to our experience with wind models of O stars this
expression is valid in the wind region close to the star with
$r\lesssim1.2 R_*$ within
the accuracy of
$20\%$.
This is
the
region of rapid wind acceleration where the stellar wind attains
velocities of
the order of
$100\,\text{km}\,\text{s}^{-1}$.
Note, however, that obtained results do 
not significantly depend on the detailed form of the velocity law.
The density structure is obtained from the equation of continuity.
Wind temperature in these models was equal to $T_\text{wind}$ and
electron density was consistently calculated from hydrogen and helium
ionization balance.
For each of these stars with different effective temperatures three
models with different values of mass-loss rates were calculated, namely 
$10^{-6}\Msun\,\text{year}^{-1}$,
$10^{-7}\Msun\,\text{year}^{-1}$,
$10^{-8}\Msun\,\text{year}^{-1}$.

\begin{figure}[t]
\centering
\resizebox{\hsize}{!}{\includegraphics{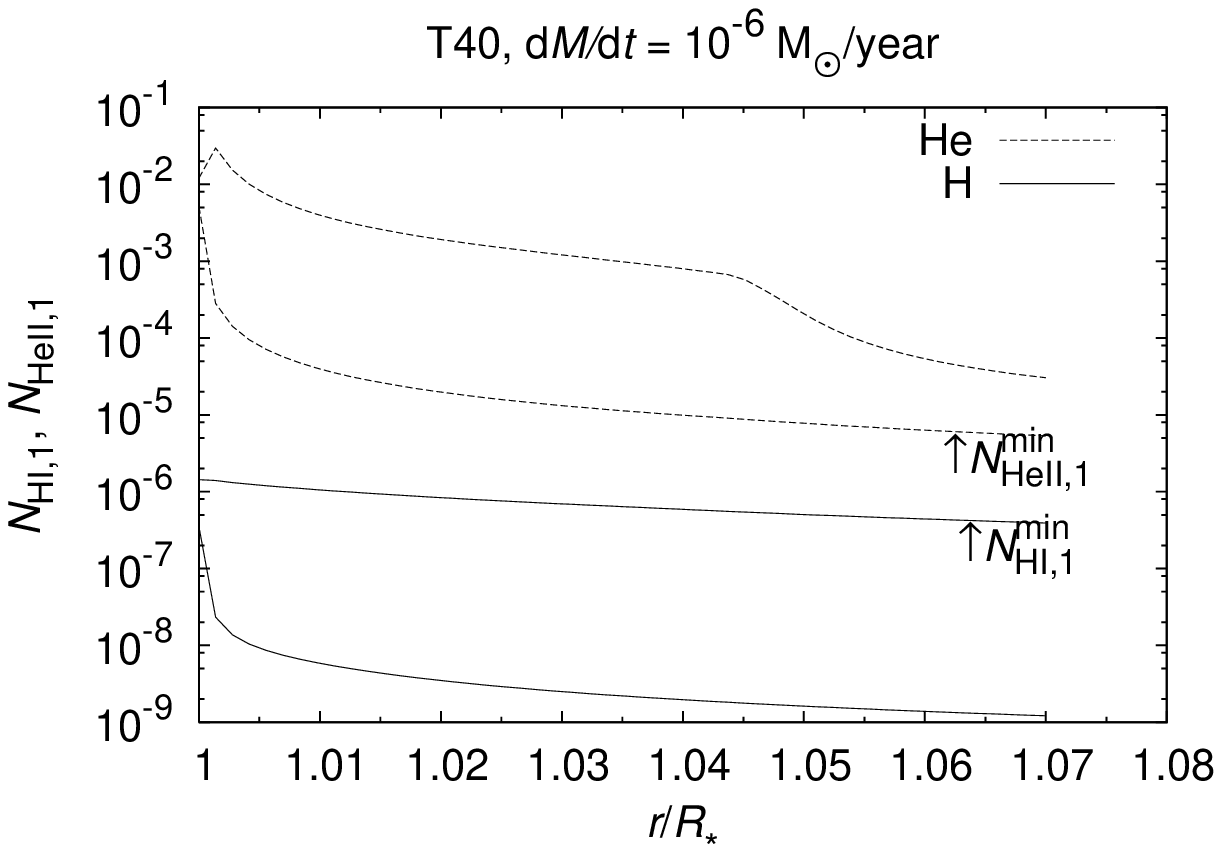}}
\resizebox{\hsize}{!}{\includegraphics{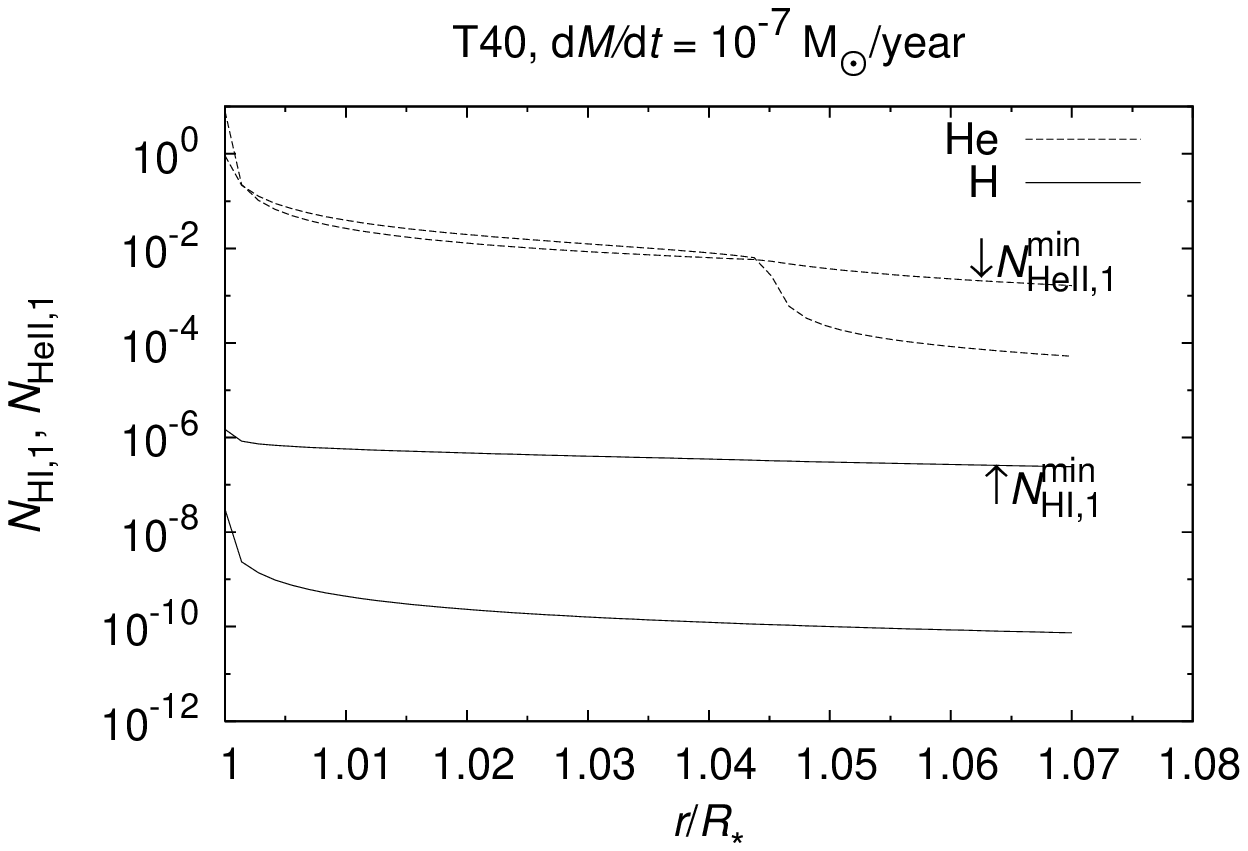}}
\resizebox{\hsize}{!}{\includegraphics{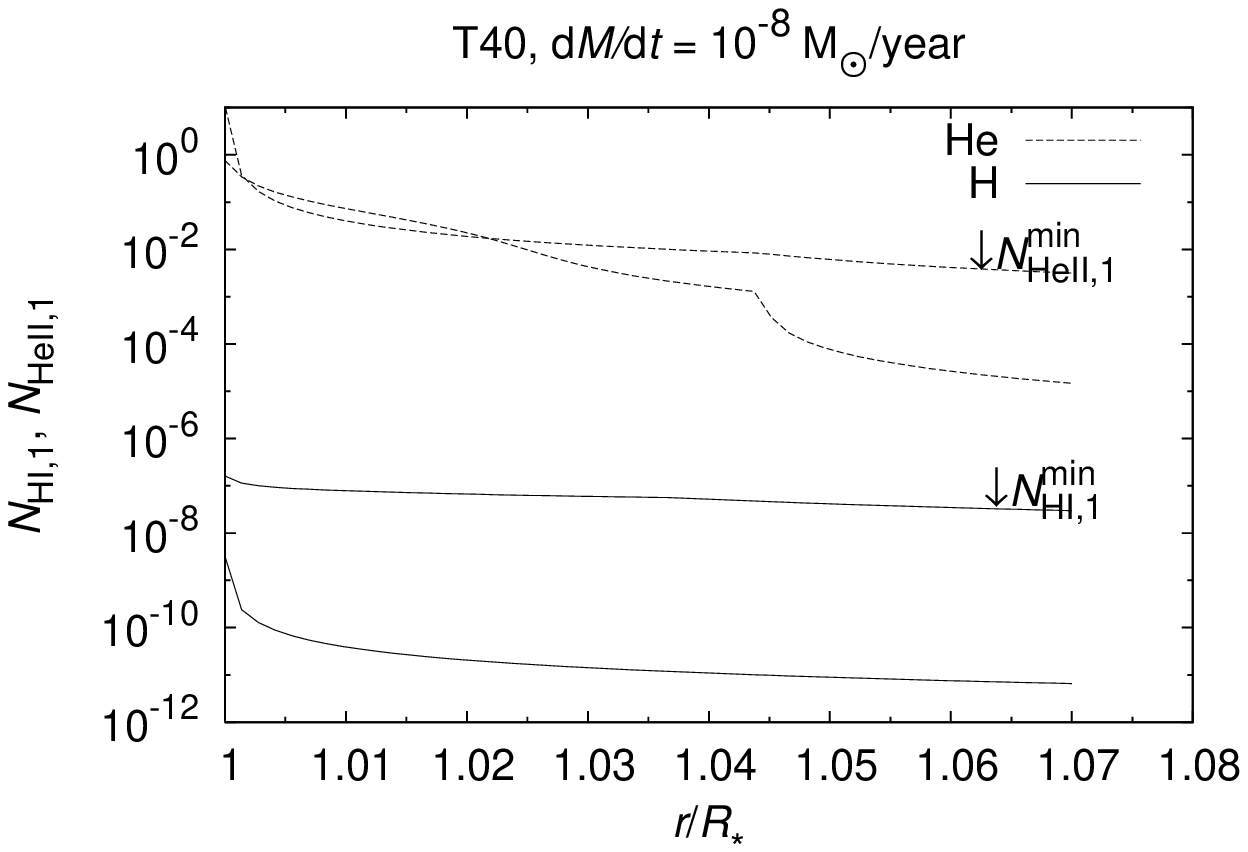}}
\caption{Comparison of minimal occupation numbers
$N_{\text{\ion{H}{i}},1}^{\min}$, $N_{\text{\ion{He}{ii}},1}^{\min}$
required to launch the wind and actual occupation numbers
$N_{\text{\ion{H}{i}},1}$ and $N_{\text{\ion{He}{ii}},1}$
(shown by unlabeled curves)
for model T40 and different mass-loss rates
of $10^{-6}\Msun\,\text{year}^{-1}$,
$10^{-7}\Msun\,\text{year}^{-1}$, and
$10^{-8}\Msun\,\text{year}^{-1}$.}
\label{obsamint40}
\end{figure}

\begin{figure}[t]
\centering
\resizebox{\hsize}{!}{\includegraphics{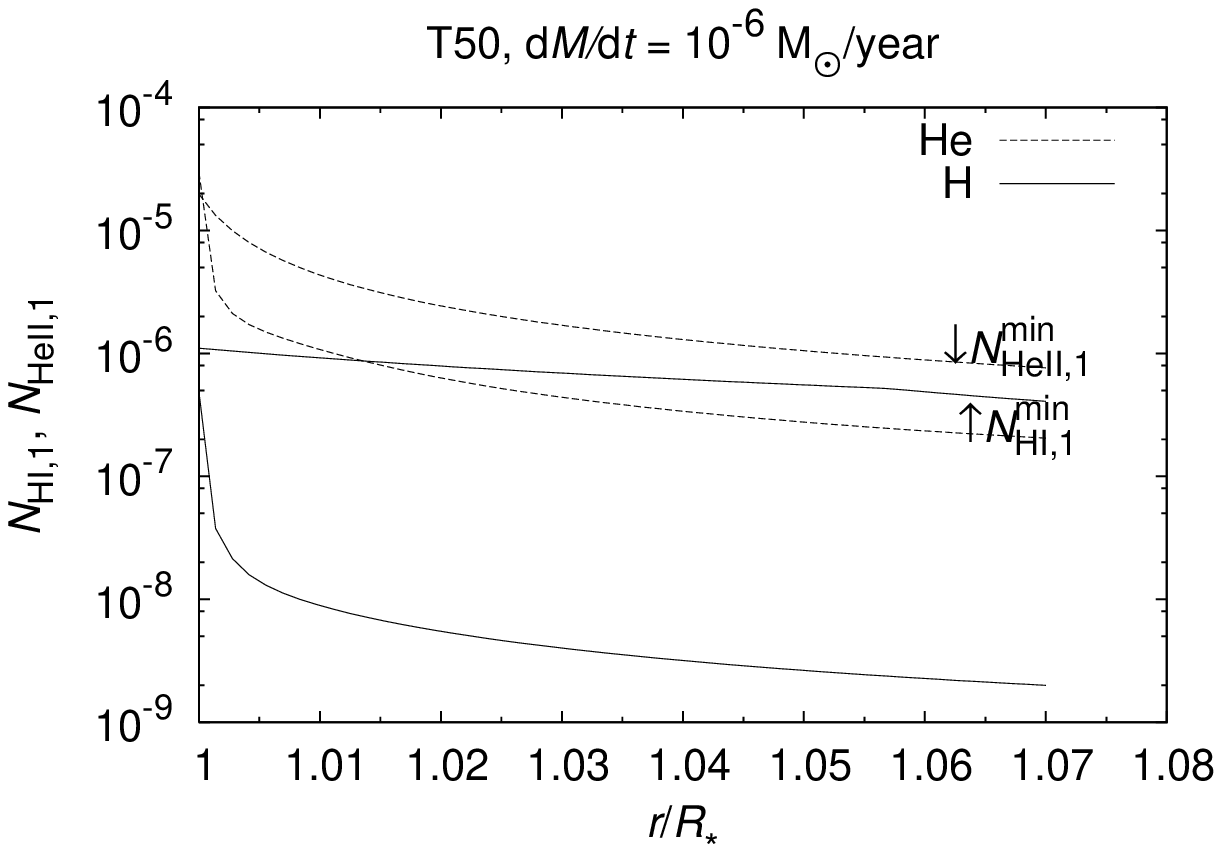}}
\resizebox{\hsize}{!}{\includegraphics{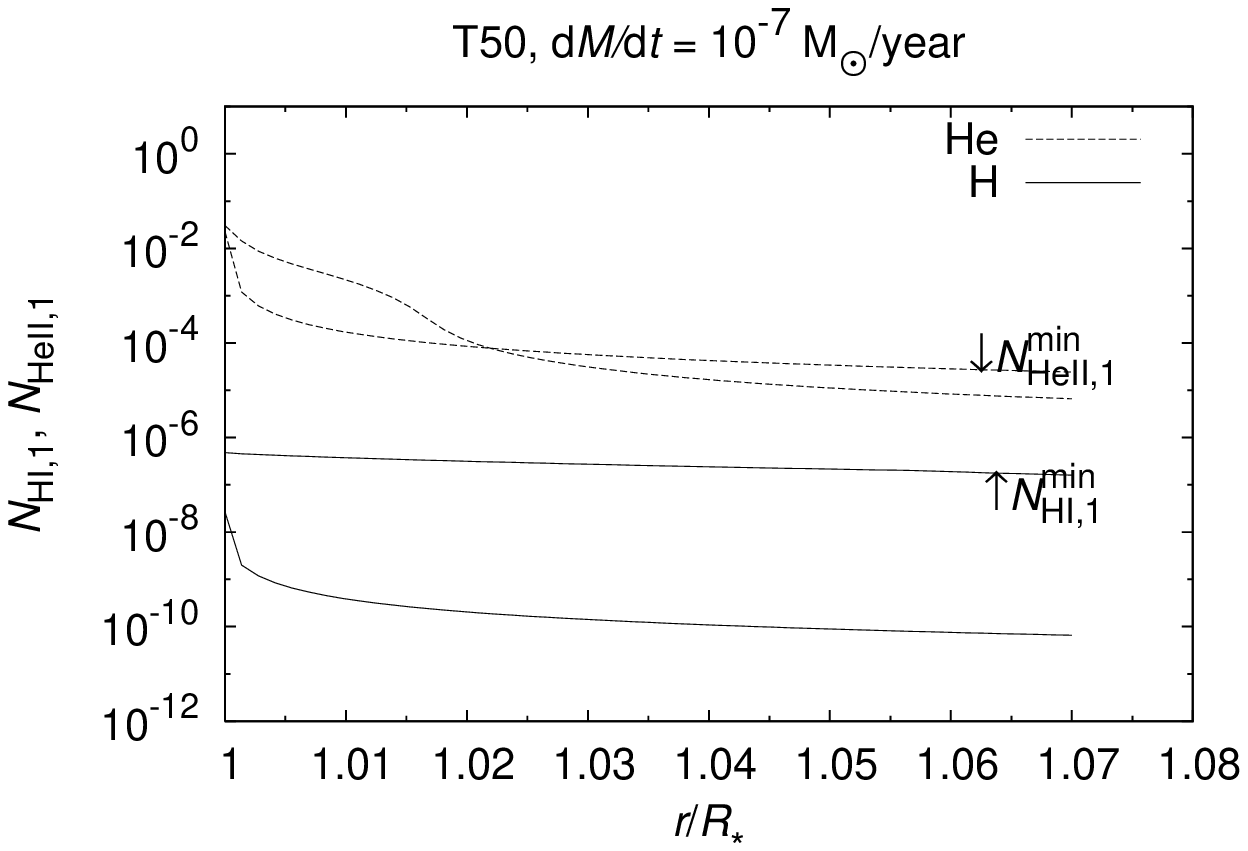}}
\resizebox{\hsize}{!}{\includegraphics{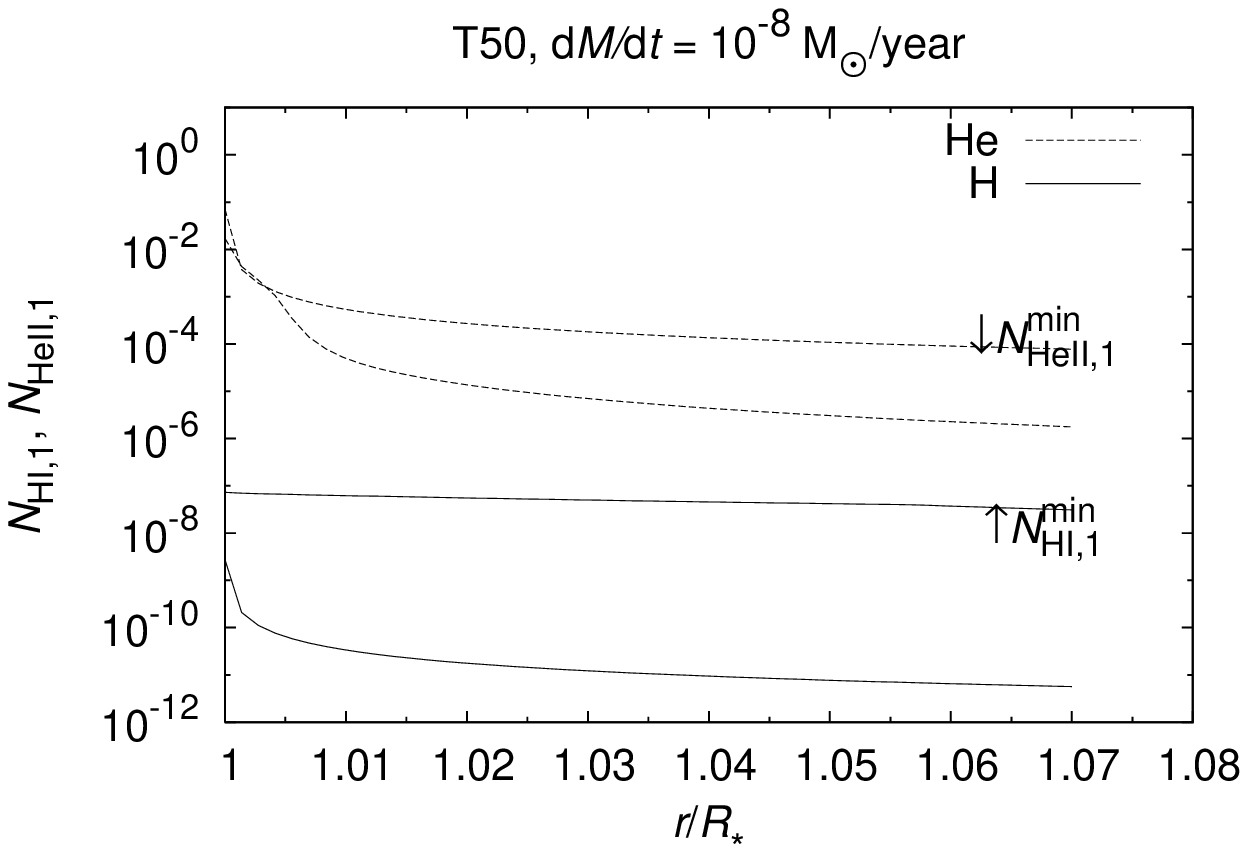}}
\caption{The same as Fig.\,\ref{obsamint40}, however for model T50.}
\label{obsamint50}
\end{figure}

\begin{figure}[t]
\centering
\resizebox{\hsize}{!}{\includegraphics{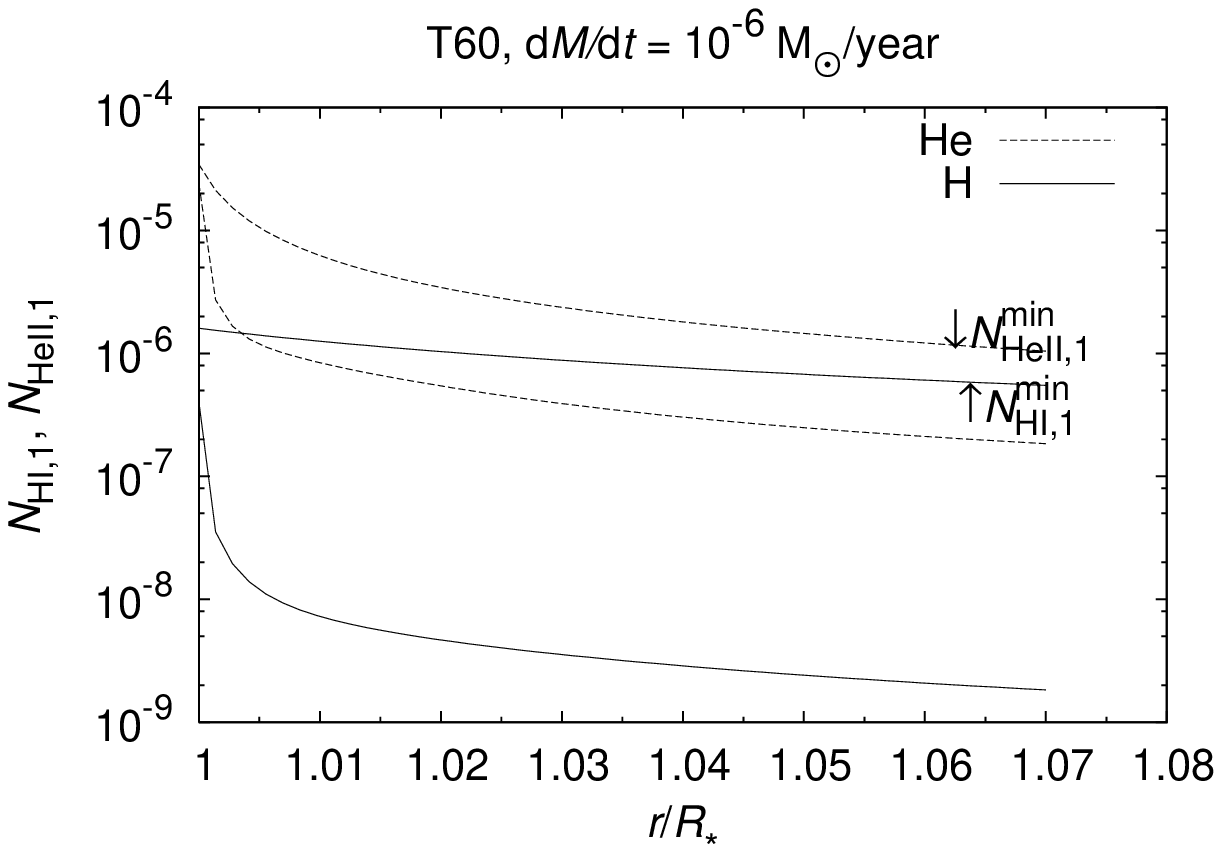}}
\resizebox{\hsize}{!}{\includegraphics{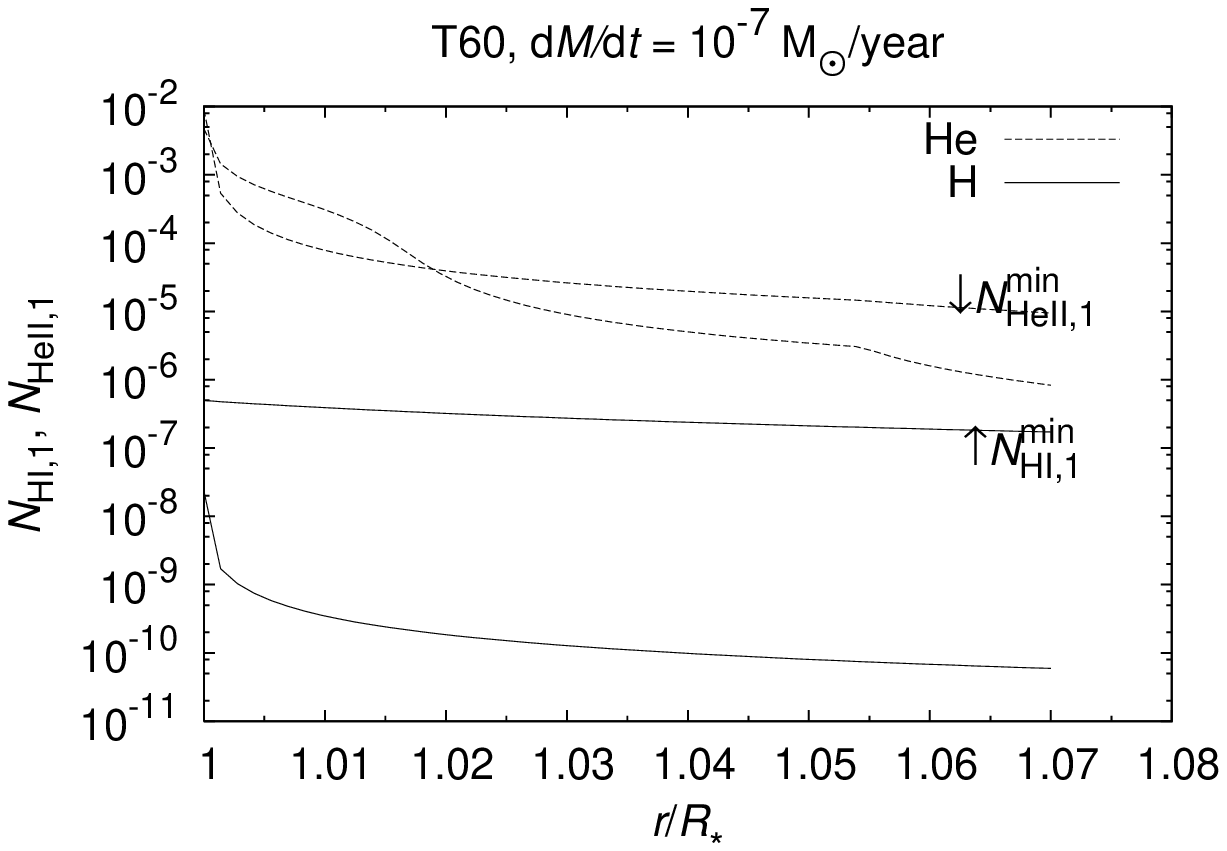}}
\resizebox{\hsize}{!}{\includegraphics{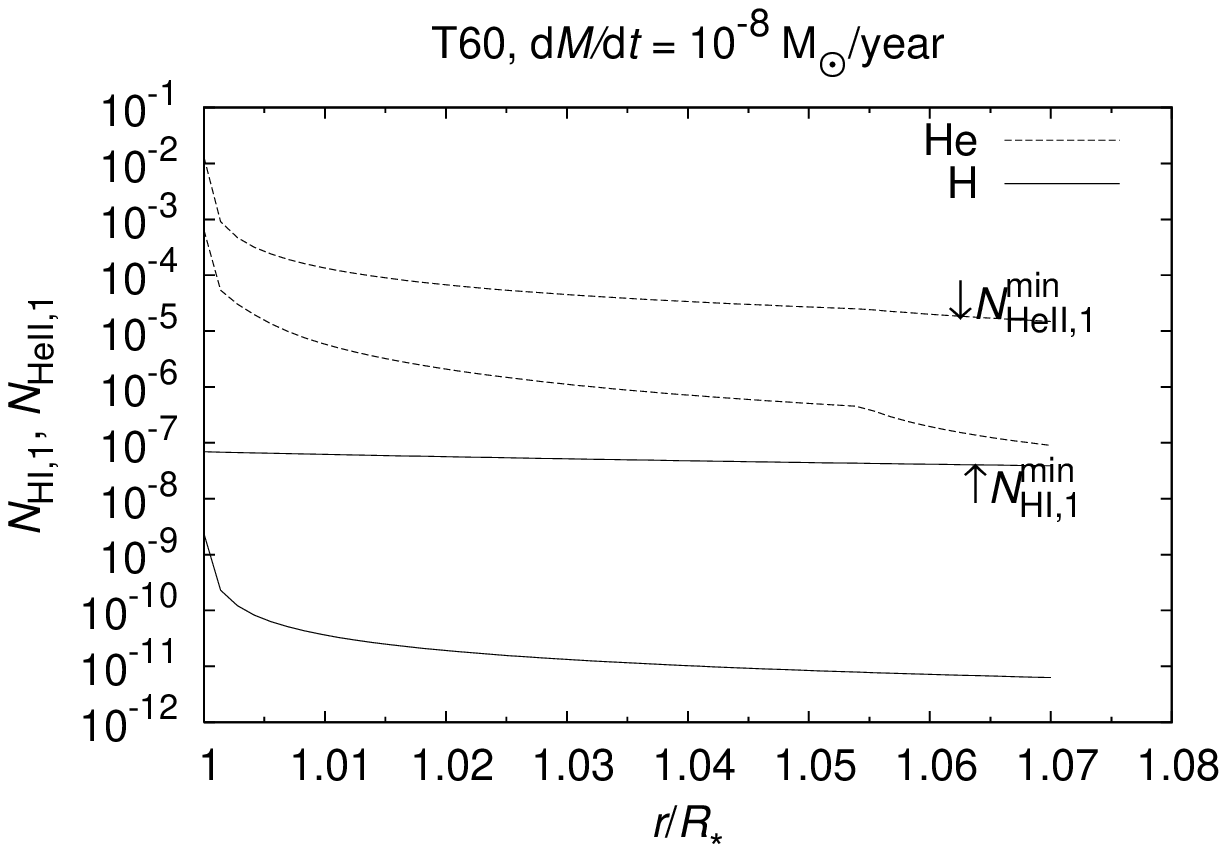}}
\caption{The same as Fig.\,\ref{obsamint40}, however for model T60.}
\label{obsamint60}
\end{figure}

Comparison of model number densities and minimum number densities
necessary to launch the wind is given in Figs.\,\ref{obsamint40} 
-- \ref{obsamint60}.
Near the stellar surface number densities of ground levels of
hydrogen-like ions are relatively high and they 
decrease with radius due to increasing ionization.
The ionization is higher in the outer regions due to both
lower density and higher radiative flux at frequencies outside the line
core
(due to the 
Doppler
effect the line frequency shifts from the saturated
core of
photospheric
line-profile, consequently
larger flux
absorbed
in the lines increases populations of higher levels, for
which the ionization rate is much higher than for lower levels).
Minimum number densities also decrease.
This is caused by increasing stellar flux as the wind lines move out
from the centre of photospheric line due to the Doppler effect.

The relative number density of the ground level of \ion{H}{i} is very
low, mostly much lower than the minimum number density necessary to
launch the stellar wind by \ion{H}{i} lines.
Thus, we conclude that
\emph{
\ion{H}{i} lines are not important for
line-driven winds of first stars.
}
On the other hand, for specific
regions of some models the actual number density of ground level
of \ion{He}{ii} is higher than the
minimum number density necessary to launch the stellar wind.
In these regions the total radiative force may be higher than the
gravity force and
the sufficient condition for launching
the
stellar wind may be fulfilled there.
%


\subsection{Maximum radiative force}

\begin{figure}[t]
\centering
\resizebox{\hsize}{!}{\includegraphics{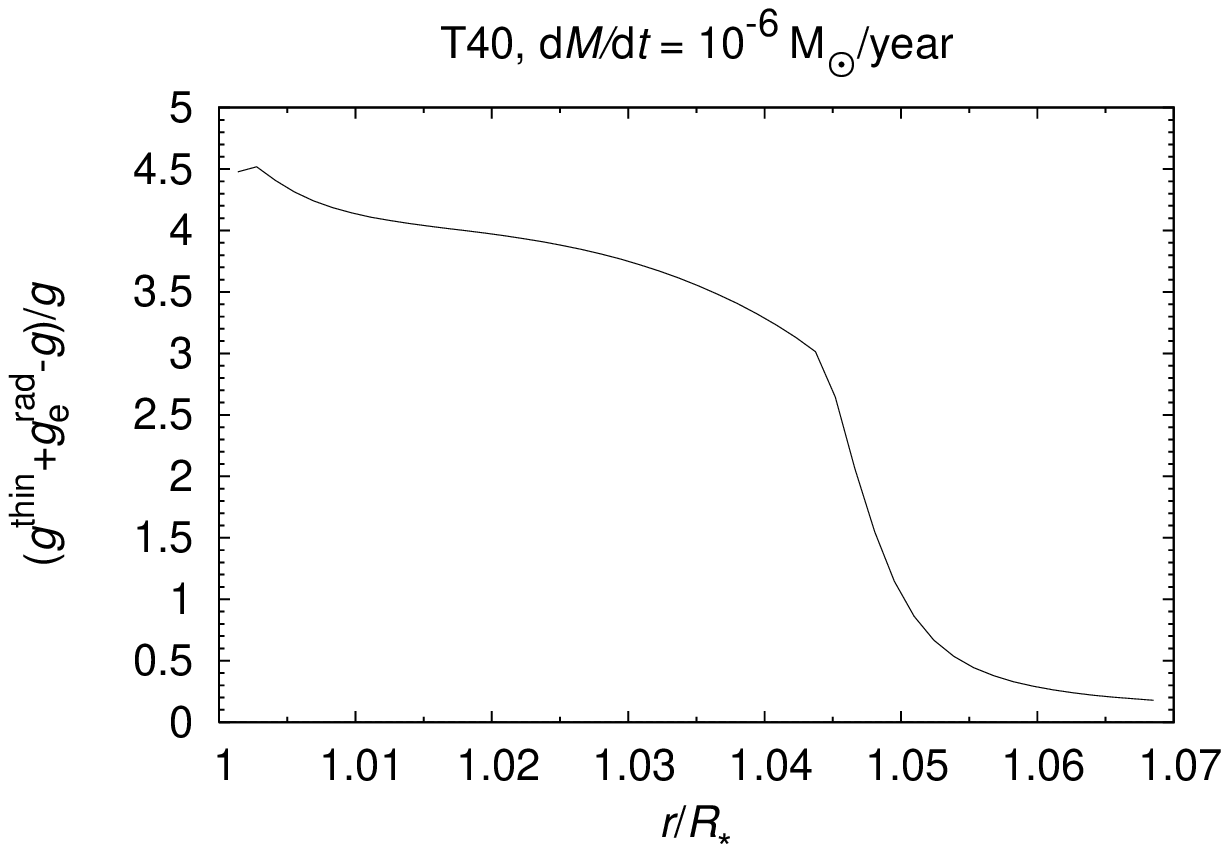}}
\resizebox{\hsize}{!}{\includegraphics{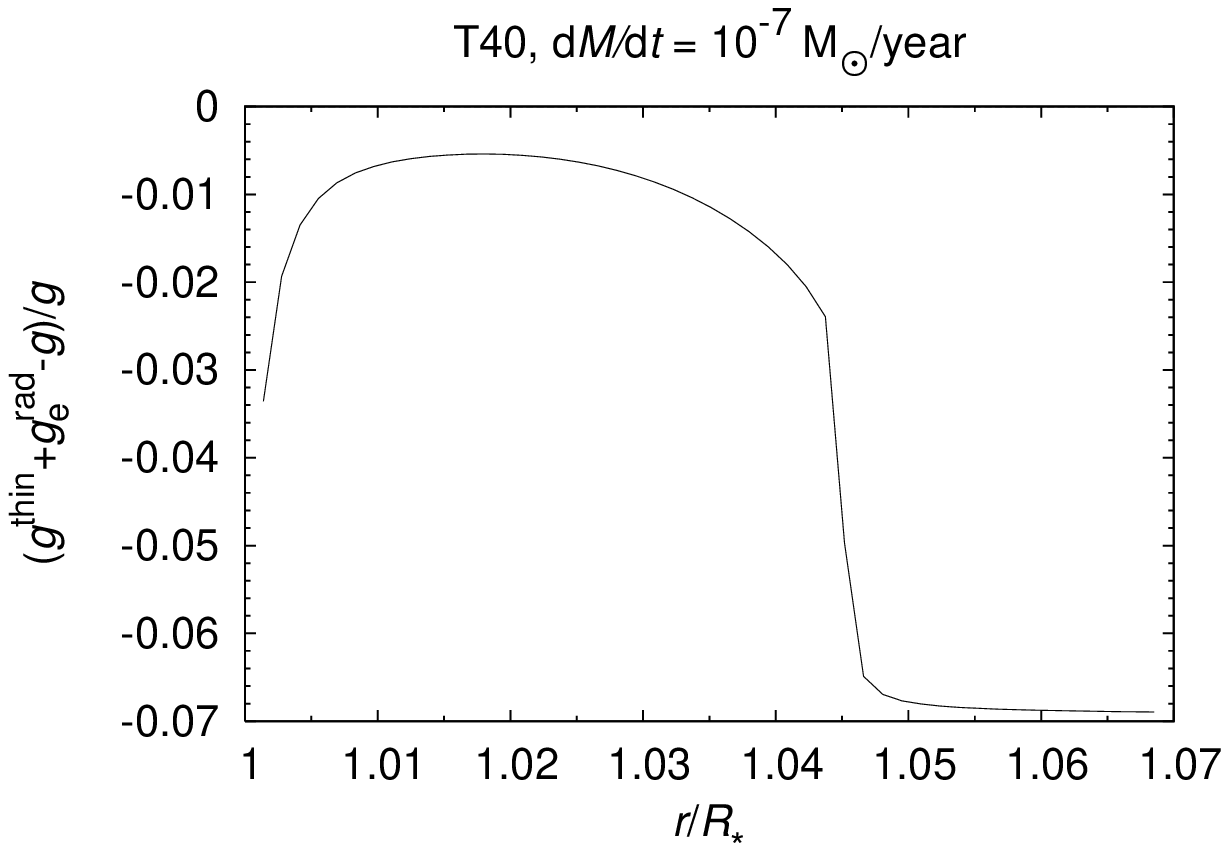}}
\resizebox{\hsize}{!}{\includegraphics{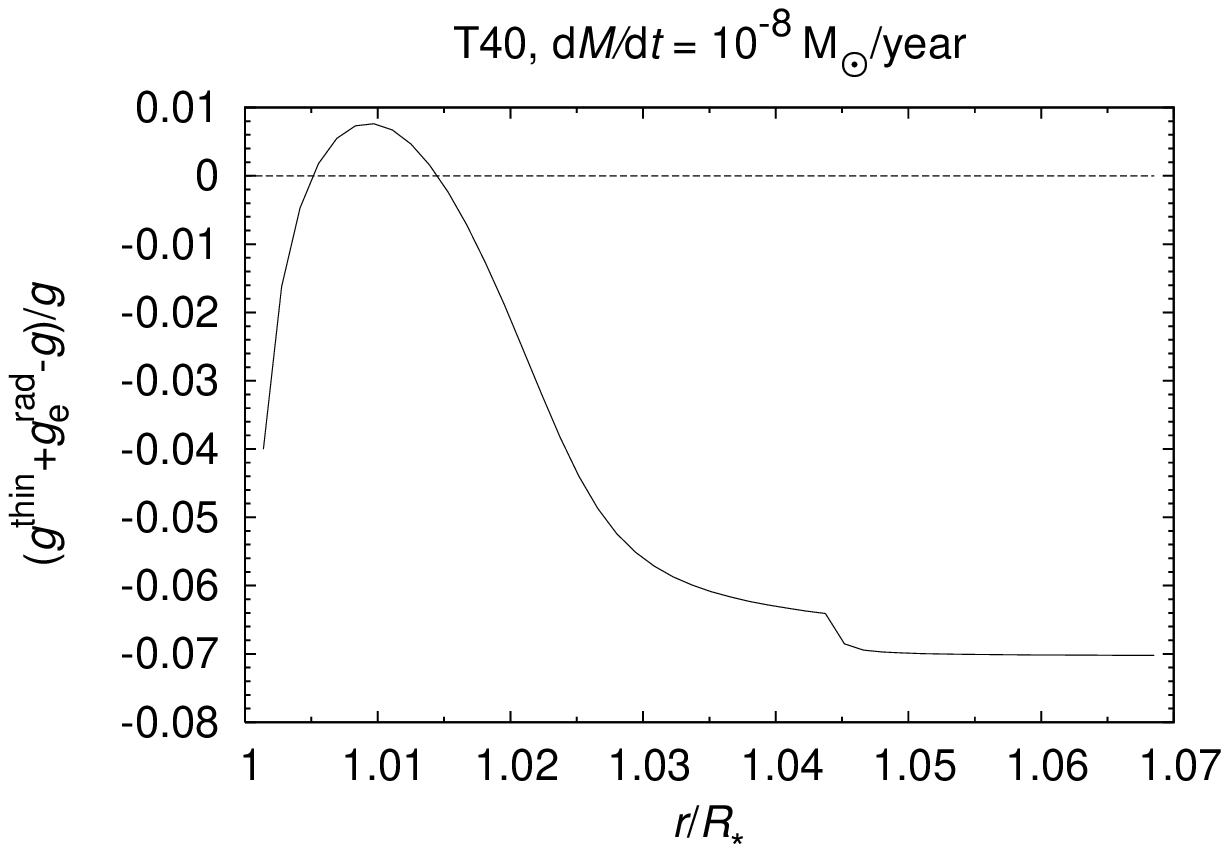}}
\caption{The plot of the relative difference
$\zav{g^\mathrm{thin}+g_\mathrm{e}^\mathrm{rad}-g}/g$ for model T40.
Note that the points where this difference is zero match the points
where the actual occupation number $N_{\text{\ion{He}{ii}},1}$ is equal
to $N_{\text{\ion{He}{ii}},1}^{\min}$ relatively well.
Thus, the simple conditions Eqs.\,\eqref{nejmennh} and \eqref{nejmennhe}
for the minimum occupation number necessary to launch the stellar wind
are able to predict relatively correctly the regions where the total
radiative force is greater than gravity.}
\label{rozmint40}
\end{figure}

\begin{figure}[t]
\centering
\resizebox{\hsize}{!}{\includegraphics{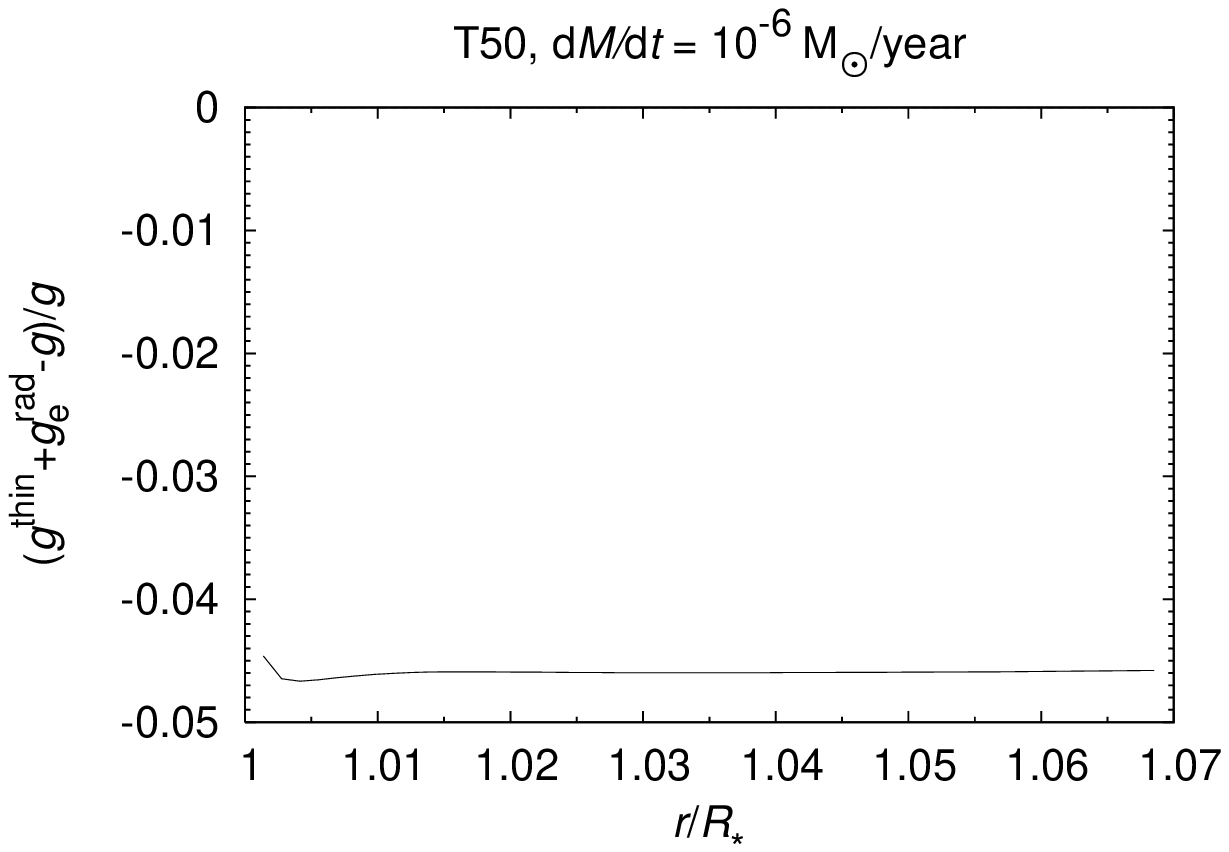}}
\resizebox{\hsize}{!}{\includegraphics{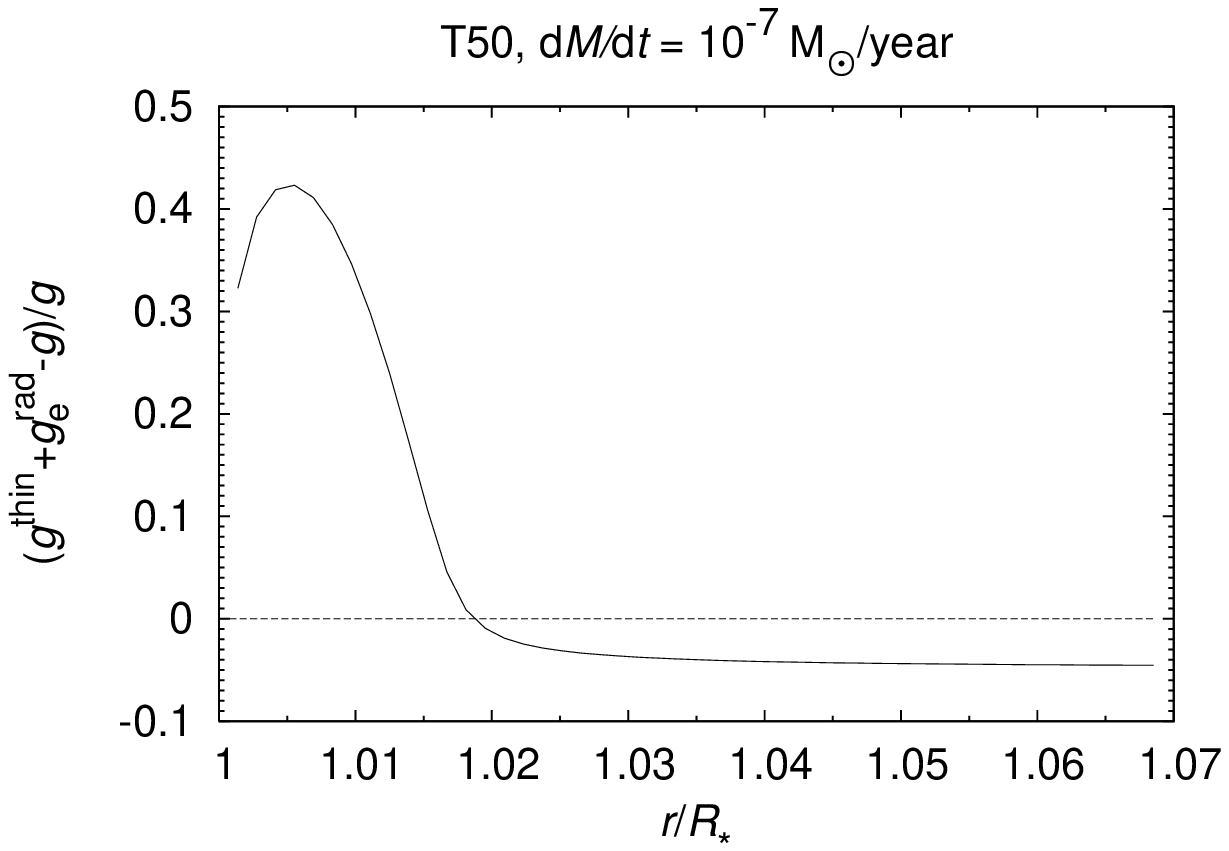}}
\resizebox{\hsize}{!}{\includegraphics{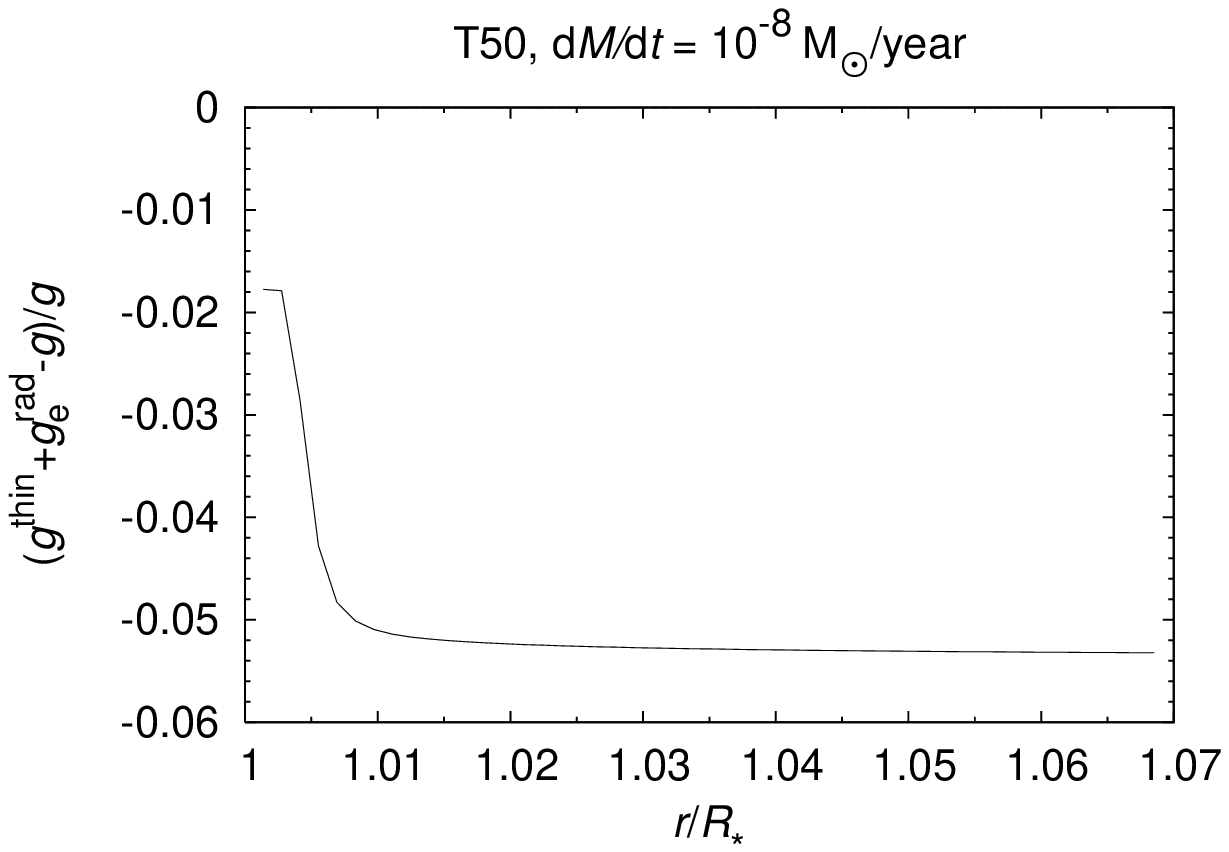}}
\caption{The same as Fig.\,\ref{rozmint40}, however for model T50.}
\label{rozmint50}
\end{figure}

\begin{figure}[t]
\centering
\resizebox{\hsize}{!}{\includegraphics{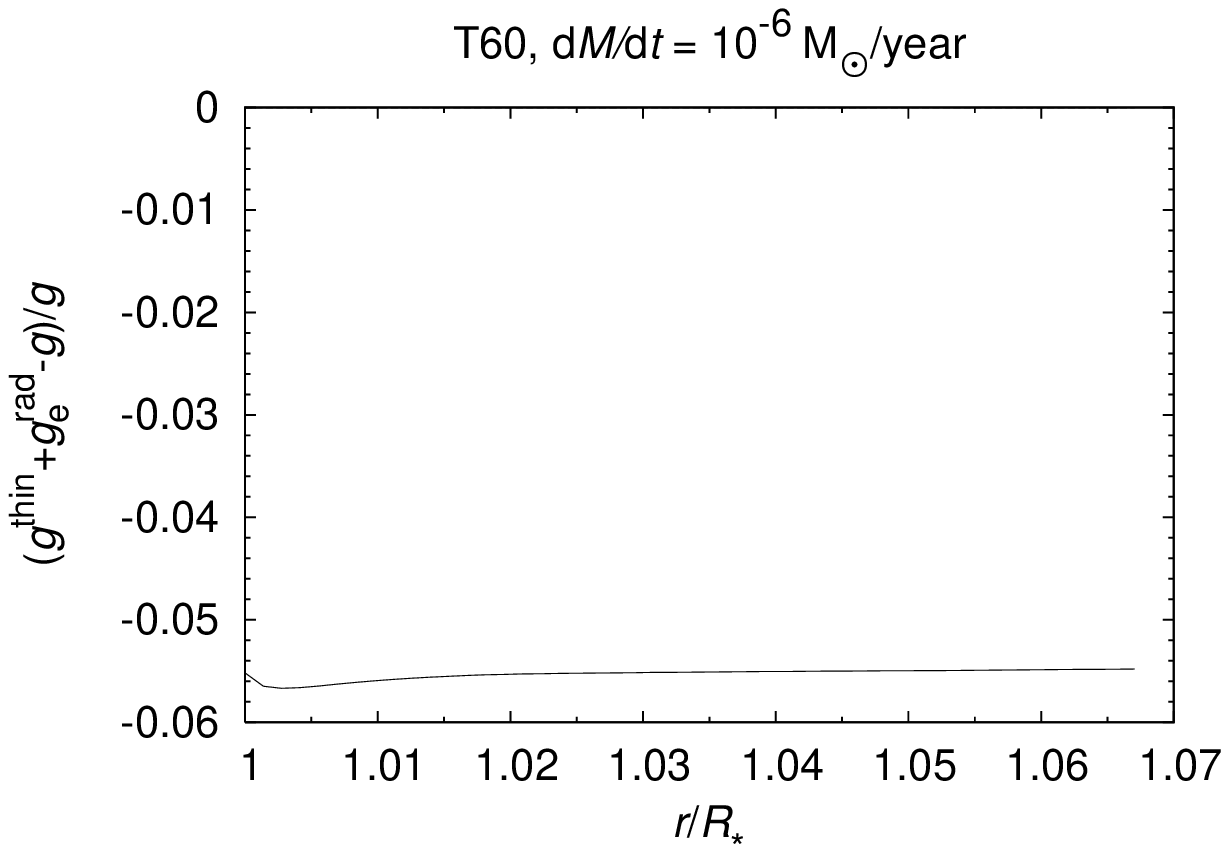}}
\resizebox{\hsize}{!}{\includegraphics{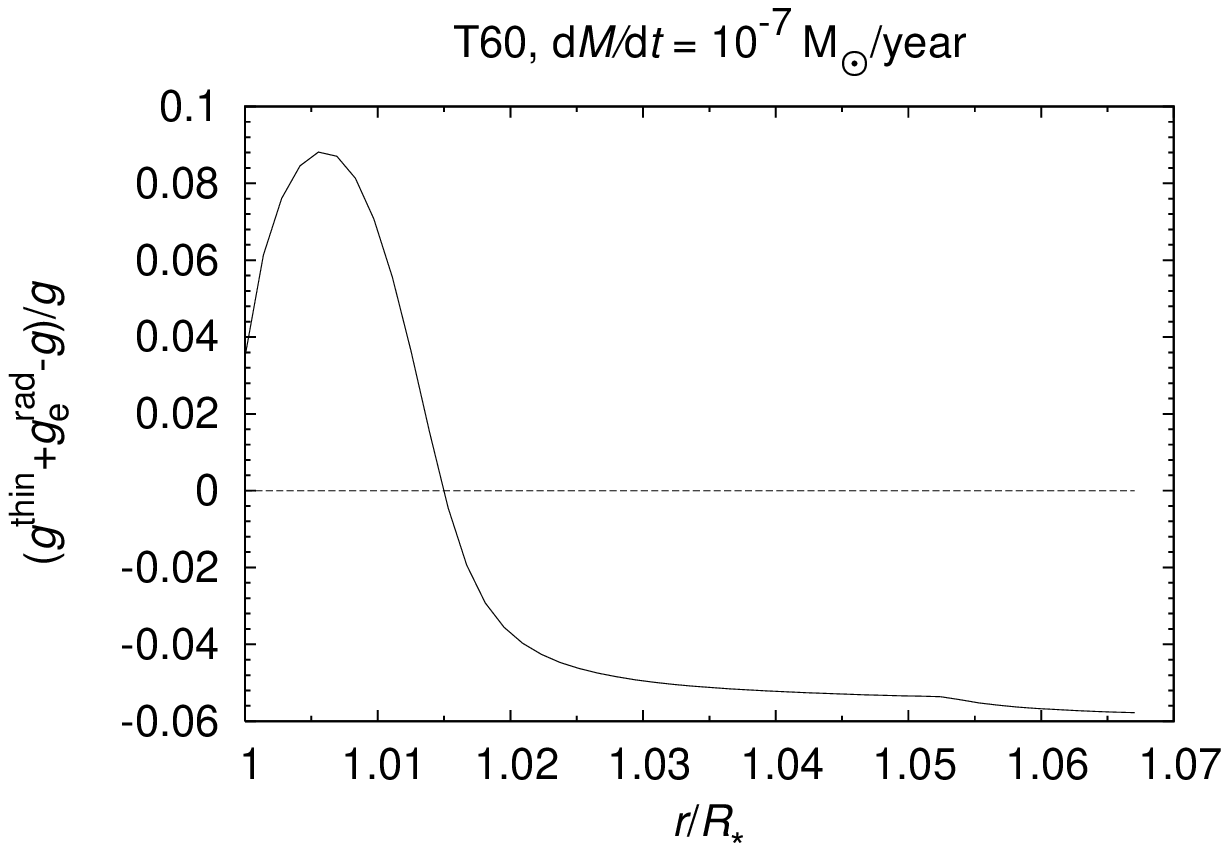}}
\resizebox{\hsize}{!}{\includegraphics{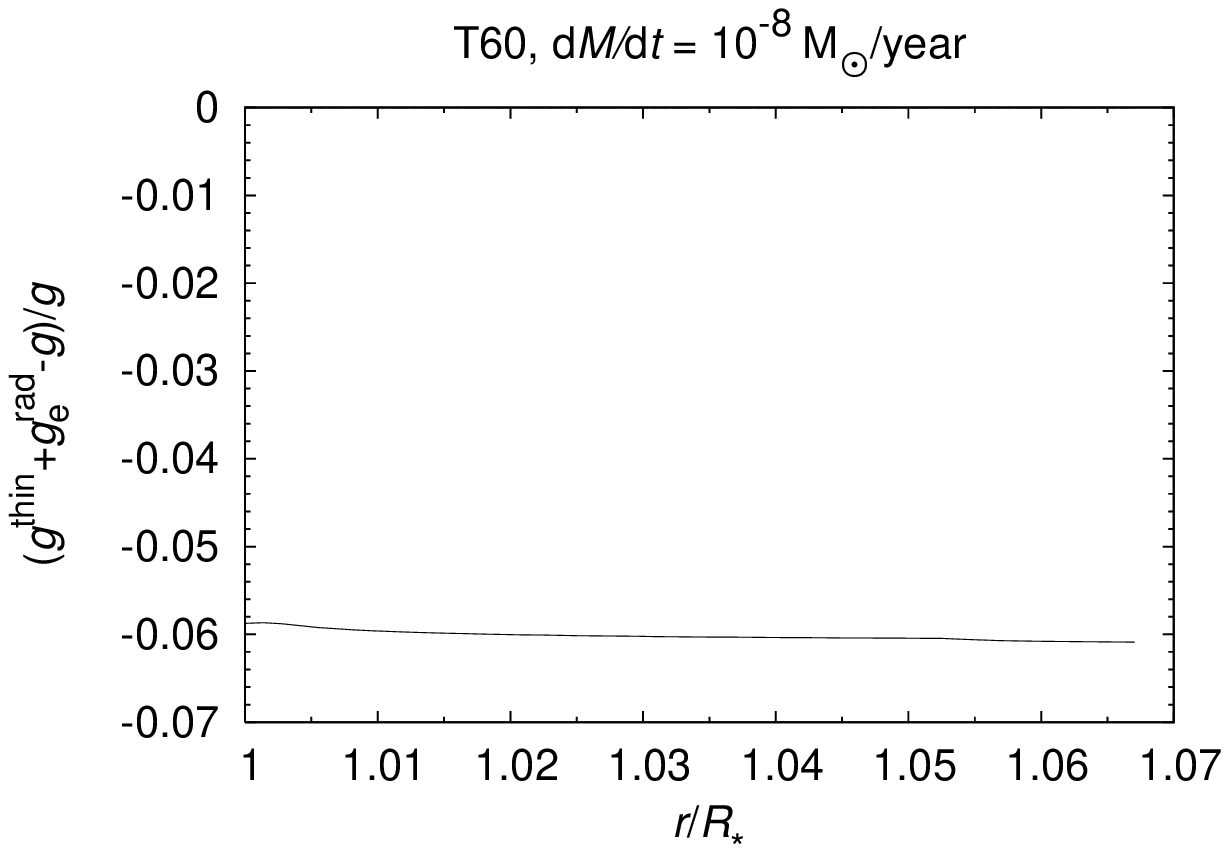}}
\caption{The same as Fig.\,\ref{rozmint40}, however for model T60.}
\label{rozmint60}
\end{figure}

\zm{The}
preceding simple analysis can be extended using the detailed calculation
of the radiative force.
For stars in Tab.\,\ref{porovioni} we have calculated the relative
difference
$\zav{g^\mathrm{thin}+g_\mathrm{e}^\mathrm{rad}-g}/g$
(i.e. the net acceleration in terms of gravity acceleration)
as a function of radius.
Corresponding graphs are plotted in Figs.\,\ref{rozmint40}~--~\ref{rozmint60}.

We came to the same conclusion as in the previous paragraph.
The maximum possible radiative force is greater than the gravity in some
regions of some models.
However,
the functional dependence of the difference
$g^\mathrm{thin}+g_\mathrm{e}^\mathrm{rad}-g$ 
in most cases does not have a correct radial dependence
necessary to launch the stellar stellar wind,
i.e. the total radiative force in these models is higher
than the gravity near the star and lower than the gravity in the outer regions.


Moreover, we have to keep in mind that the optically thin radiative
force is maximum for a given set of lines and clearly the actual
radiative force will be lower due to the self-shadowing effect.
Moreover, for many of the models the total radiative force is not
greater than the gravity.
Thus, we conclude that
\emph{
winds of zero metallicity stars driven purely by
the optically thin lines are unlikely.
}
However, these results shall be later
firmed up by a more detailed models with
the possibility that some lines are optically thick.

Inspecting Figs.\,
\ref{obsamint40}~--~\ref{rozmint60} 
note that 
Eqs.\,\eqref{nejmennh} and \eqref{nejmennhe} are able to reliably 
estimate the minimum occupation number necessary to launch a line-driven
stellar wind because
they
correctly
predict the location of region where the optically thin
radiative force is greater than gravity.

%

\section{Wind with optically thick lines}

Our final test of the possibility of the existence of line-driven
stellar wind of first stars should include the natural possibility that
some lines are optically thick. Thus, we now turn our attention to 
more realistic case of
the stellar wind driven partially by the optically thick lines.
This is a much more complicated task than the wind driven purely by
optically thin lines since now we cannot use any simple criterion to
decide whether the wind is possible or not.
We have to use detailed modelling.
Moreover, the nonlinear character of wind equations (especially
equations of statistical equilibrium) does not guarantee that the
obtained solutions are unique.
This is especially the case of zero metallicity stars when the wind may
be driven just by one ionization state.
Simply, for some ionization structures the wind may be
possible and for some other not.

To make this problem more tractable we calculate radiative force in the
Sobolev approximation for the same models, for which we studied
optically thin radiative force in Sect.\,\ref{tenkovit}.
The radiative force in the Sobolev approximation is given by the sum of
partial radiative forces due to the individual lines (Castor \cite{cassob},
Abbott \cite{abpar}) 
\begin{equation}
\label{zarsil}
g^{\rad}=\frac{8\pi}{\rho c^2}\frac{\vr}{r}
 \sum_{\mathrm{lines}}\nu_{ij} H_c\int_{\mu_c}^{1}\mu\zav{1+\sigma\mu^2}
 \zav{1-e^{-\tau_\mu}}\,\text{d}\mu,
\end{equation}
where
$\rho$ is the wind density, $\vr$ is the radial velocity,
$H_c$ is the stellar flux,
$\mu_c=\sqrt{1-R_*^2/r^2}$ is the direction cosine of the photospheric
limb, $R_*$ is stellar radius, $\sigma={\de\ln \vr}/{\de\ln r} -1$ and
the Sobolev optical depth $\tau_\mu$ is
\begin{equation}
\label{tau}
\tau_\mu=\frac{\pi
e^2}{m_\mathrm{e}\nu_{ij}}\zav{\frac{n_i}{g_i}-\frac{n_j}{g_j}}
g_if_{ij}
\frac{r}{\vr\zav{1+\sigma\mu^2}}.
\end{equation}

If some of the lines are optically thick, then the actual radiative
acceleration
given in the Sobolev approximation Eq.\,\eqref{zarsil} is lower
than the maximum radiative 
acceleration 
\eqref{silazarmax}.
This is caused by the effect of self-shadowing, i.e. that some part of
the stellar flux which is supposed to accelerate the wind in a line was
already absorbed by a given line in the region closer to the star.
Thus, here we present only results for models for which optically thin
radiative force can drive a stellar wind, since it is clear that there
will not be any wind in the opposite case. 

\begin{figure*}[t]
\centering
\resizebox{0.49\hsize}{!}{\includegraphics{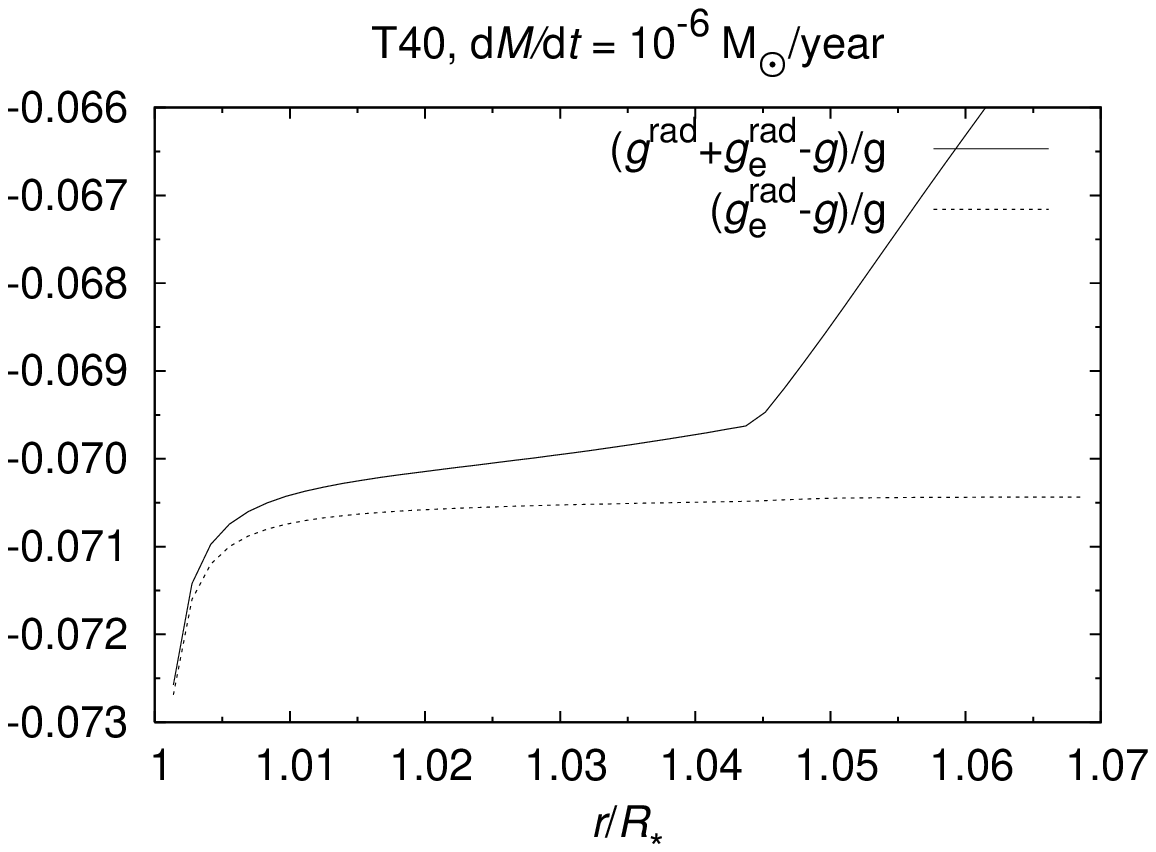}}
\resizebox{0.49\hsize}{!}{\includegraphics{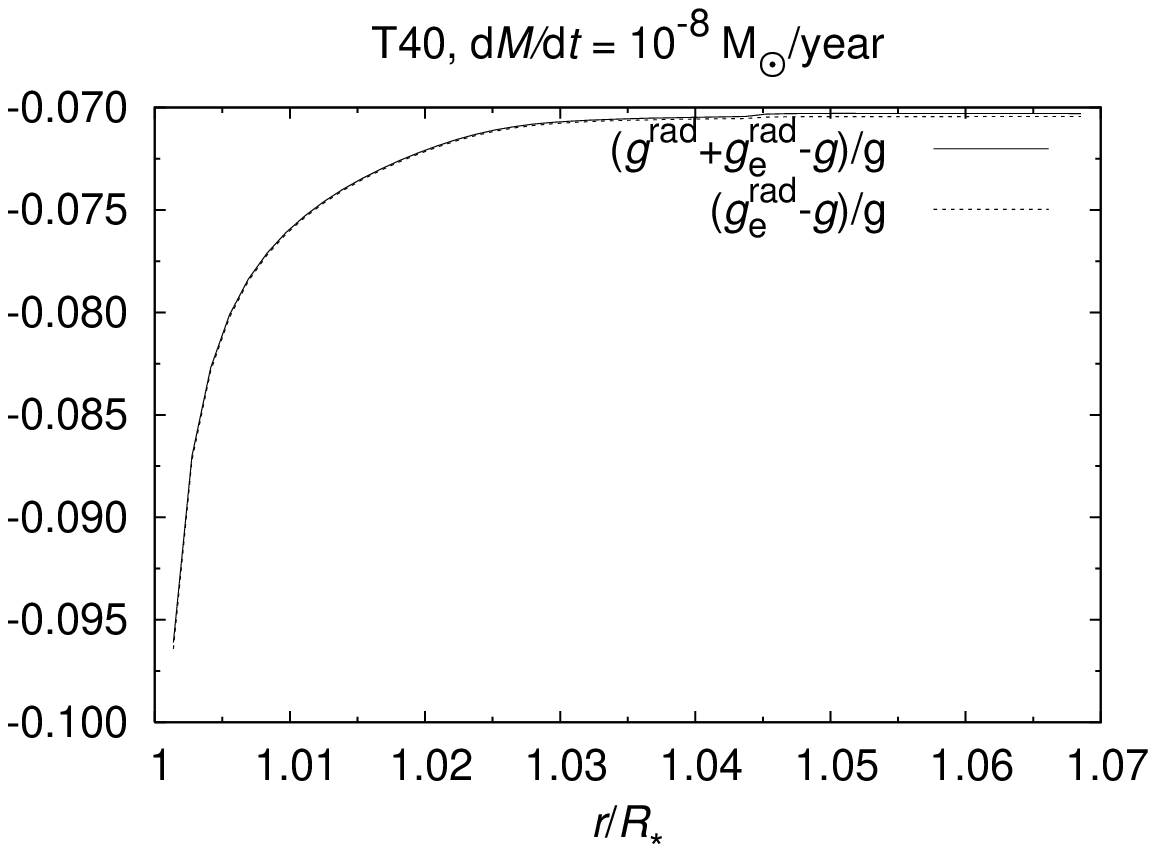}}
\resizebox{0.49\hsize}{!}{\includegraphics{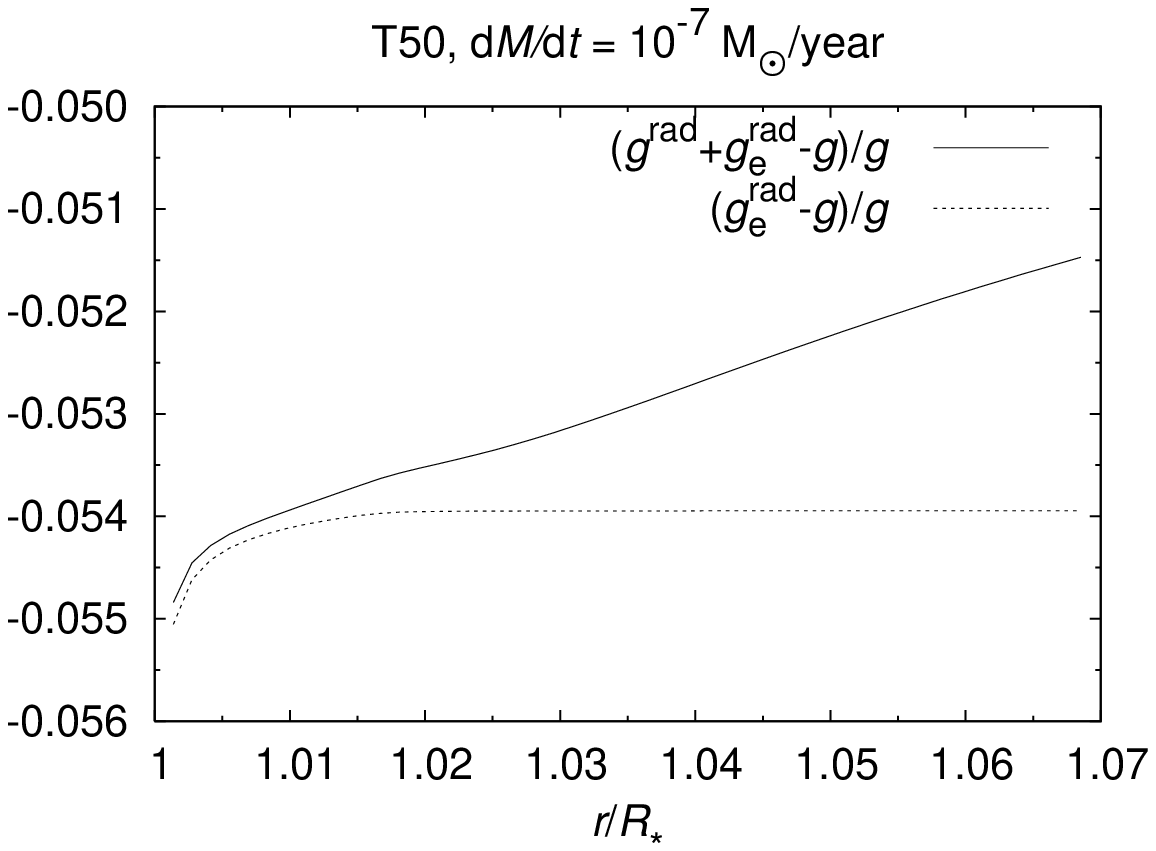}}
\resizebox{0.49\hsize}{!}{\includegraphics{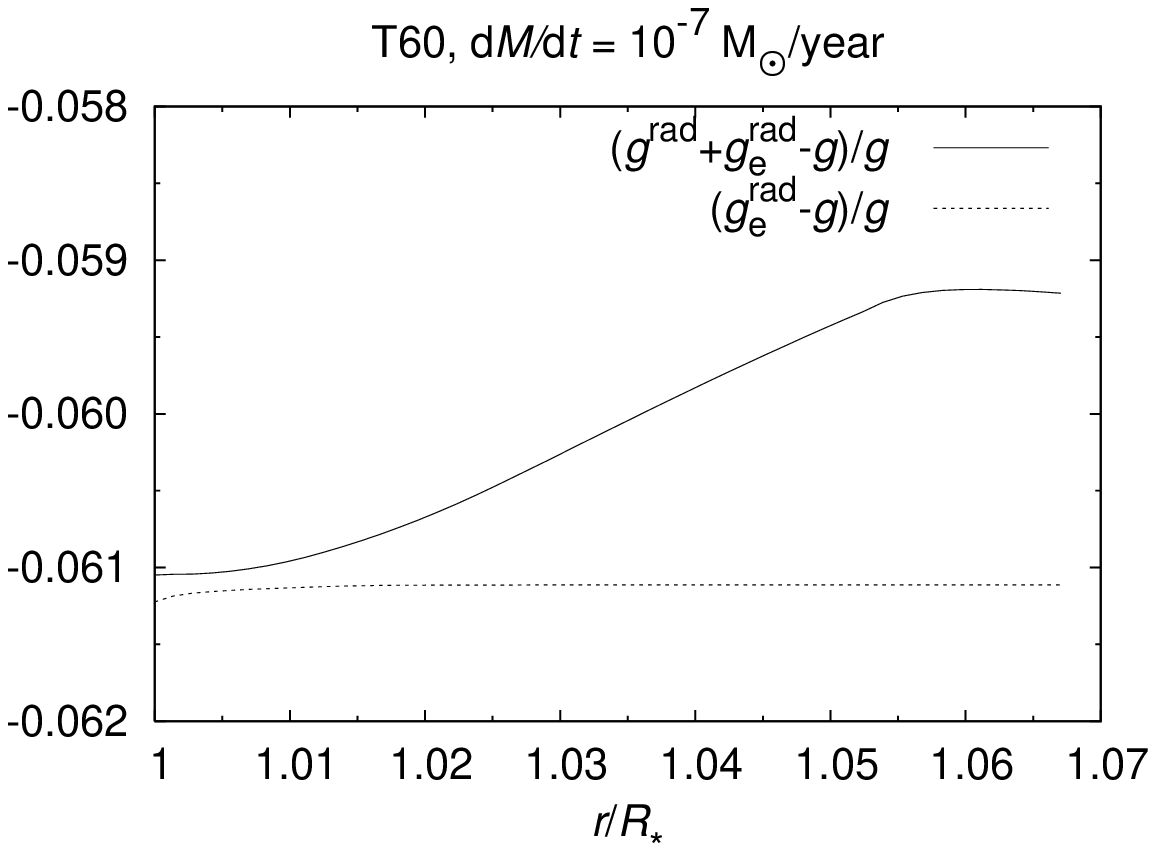}}
%
\caption{The plot of the 
net
radiative acceleration
$g^\mathrm{rad}+g_\mathrm{e}^\mathrm{rad}
-g
$ and the
net
radiative acceleration of the whole gas due to free electrons only
$g_\mathrm{e}^\mathrm{rad}
-g
$ relative to the gravitational acceleration
$g$.
Clearly, the contribution of the radiative acceleration due to the line
transitions $g^\mathrm{rad}$ calculated after
Eq.\,\eqref{zarsil}
is much smaller than the gravity acceleration and is not able to drive
a stellar wind.}
\label{zarsilobr}
\end{figure*}

The radiative force calculated using Eq.\,\eqref{zarsil} in the selected
models is given in Fig.\,\ref{zarsilobr}.
Clearly, since most of the 
strongest
lines are optically thick, the
radiative force due to these lines is much lower than the optically thin
radiative force. 
More realistic radiative force given by Eq.\,\eqref{zarsil} is generally
by several orders of magnitude lower than the gravity force.
Thus it is unlikely that some change of wind density or velocity may
give significantly stronger radiative force which would drive a stellar
wind.
This is also supported by our models, because we were not able to find
sufficiently large radiative force for any reasonable wind models.
\emph{
We conclude that it is unlikely that stationary massive
\element{H}-\element{He} stars have any \element{H}-\element{He}
stellar wind driven by the line transitions.
}

We have also calculated similar models for 
other stars from
Kudritzki (\cite{kudmet}) list with lower masses and lower luminosities
(with parameters given in Tab.~\ref{porovioni}).
These stars are not so close to the Eddington limit as stars
discussed sofar.
Consequently, the radiative force due to light scattering on free
electrons is lower for these additional stars and the minimum radiative
force necessary to launch the stellar wind is higher in this case (see
Eqs.\,\ref{nejmennh}, \ref{nejmennhe}).
Clearly, the possibility for launching of the stellar wind is lower for
stars which are not so close to the Eddington limit.




\section{Influence of continuum radiative force}

Another force which can potentially accelerate the stellar wind of first
stars is the radiative force due to bound-free and free-free
transitions.
Numerical calculation showed that the contribution of free-free
transitions to the radiative force in the circumstellar environment is
much lower than that of bound-free transitions.
Thus, we shall discuss the bound-free transitions only, although the
free-free transitions are also included in our models.
The radiative acceleration has in this case the form (Mihalas \cite{mihalas})
\begin{equation}
\label{silazarko}
g^{\text{rad,\,bf}}=\frac{4\pi}{c}\int_0^\infty \frac{\chi_\nu}{\rho}H_\nu\,
\de\nu,
\end{equation}
where $\chi_\nu$ is the opacity due to the bound-free transitions,
\begin{equation}
\label{opaciko}
\chi_\nu=\sum_i\hzav{n_i-n_{k}\zav{\frac{n_i}{n_{k}}}^*
\text{e}^{-\frac{h\nu}{kT}}}\alpha_{ik}(\nu),
\end{equation}
where summation goes over all levels $i$ for which bound-free
transitions are accounted for
and
the
asterisk refers to
the
LTE value.

Let us first briefly discuss an interesting possibility that the stellar
wind is driven purely by the bound-free absorption.
Combining Eqs.\,\eqref{silazarko} and \eqref{opaciko}, the radiative
acceleration is given by 
\begin{multline}
g^{\text{rad}}=\frac{4\pi}{c}
\sum_i
\left[\frac{n_i}{\rho} \int_{\nu_i}^\infty \alpha_{ik}(\nu) H_\nu\, \de\nu-
\right.\\*
\left.-\frac{n_{k}}{\rho}\zav{\frac{n_i}{n_{k}}}^*
\int_{\nu_i}^\infty \alpha_{ik}(\nu) H_\nu \text{e}^{-\frac{h\nu}{kT}}\, \de\nu
\right]
.
\end{multline}
As has been already mentioned in Sect.\,\ref{kaptenvit}, in order to
obtain smooth wind solution, the radiative force has to be smaller than
the gravity force in the stellar atmosphere below the critical point and
larger than the gravity force above the critical point. 
It is very difficult to achieve this in this case, since the radiative
flux $H_\nu$ decreases with radius rapidly, especially in the case when
strong bound-free absorption is present.
Moreover, the critical condition is not a sufficient one to drive the
stellar wind, since it does not tell us anything whether the radiative
force in the outer wind regions is strong enough to lift the wind
material out from the stellar gravitational potential well.
The decrease of the stellar flux with radius due to the bound-free
absorption (see also the photon tiring effect, Owocki \& Gayley
\cite{unava}) can complicate this.
Note that the situation with the line-driven wind is simpler,
%
because the lines are usually illuminated by the unattenuated
stellar flux thanks to the Doppler effect.

We included the radiative force due to the bound-free and free-free
transitions Eq.\,\eqref{silazarko} into our models.
The numerical calculations
(for stars with parameters given in Tab.\,\ref{porovioni})
showed that the radiative force due to the
bound-free and free-free transitions does not
significantly contribute to the
acceleration of stellar wind of first stars because it is typically 
at least two orders of magnitude lower than the gravity force.

Finally, we calculated also static spherically symmetric NLTE model
\element{H}-\element{He} atmospheres (for the description of the code
see Kub\'at \cite{ATAmod}), where the radiative force was
calculated without any simplification.
It turned out that all of the models with parameters from the
Table\,\ref{porovioni} are 
\zm{sub-Eddington}
i.e. the total radiative force
is always lower than gravity.
However, for the most luminous models the value of $\Gamma$
reached the value of $0.99$.
%

\section{Expulsion of individual elements by a wind}
\label{kapunikprvku}

Whereas first stars probably fail to drive \element{H}-\element{He}
outflow, there is a possibility that hydrogen or helium are expelled
separately, i.e. that there 
exists
a pure hydrogen or helium wind.
To show whether such outflow of chemically pristine elements is possible
or not, we recalculated models with the same stellar parameters
presented in the Sect.\,\ref{tenkovit}
assuming that one element may form the wind while the other not
and compared the radiative force
in 
such
chemically homogenous wind with the gravity force acting
against
such wind.
To avoid charging of the star we assume that a corresponding number of
electrons follow the escaping element making the wind electrically
neutral.
We assume that
such wind
exists
above the \element{H}-\element{He} photosphere.

Since these models are chemically homogenous, the electron density in
such models is calculated only
from the contribution of the element that remained in the wind
(i.e. either hydrogen or helium).
This difference is
crucial for the final results, at least in the case of helium.
If we assume that the helium gas consists 
of only
the isotope
\element[][4]{He}, we may say that helium has two times less electrons
per nucleon (proton or neutron) than hydrogen.
Hence, the radiative force per unit of mass 
(i.e. the radiative acceleration)
due to free electrons is lower in the case of helium.
Consequently, although the number of free electrons {\em per
atomic
nucleus} 
is higher in pure \element{He} than in \element{H}-\element{He}
models,
the total radiative force is lower.
Surprisingly, although helium
lines
significantly 
contribute
to the radiative
force in the \element{H}-\element{He} models,
the possibility of pure \element{He} outflow is lower than that of
\element{H}-\element{He}.

On the other hand, 
also
the situation in 
pure
hydrogen models
is
different.
Hydrogen plasma has higher number of free electrons per nucleon than
\element{H}-\element{He} plasma.
Consequently, \element{H}-\element{He}
stars which are close to the Eddington limit may have slightly higher
radiative force than gravity force for the same stellar parameters
(mass, radius, luminosity) in a pure hydrogen wind,
whereas the radiative force in hydrogen-helium atmosphere is (due to the
contribution of heavier helium atom) lower than gravity force.

These considerations are supported by our numerical models.
The total radiative force in helium models
is lower than gravitational force in all models considered
(for stars with parameters given in Tab.\,\ref{porovioni}).
On the other hand, for model stars which are very close to the
Eddington limit, the radiative force in purely hydrogen models
may be greater than gravity, thus potentially enabling hydrogen wind.

Condition for the existence of such hydrogen wind is relatively simple.
If the wind exists,
radiative acceleration due to the free electrons in a pure hydrogen
plasma
is
greater than the gravitational acceleration,
\begin{equation}
g^\text{rad}_\text{e,\,H}>g,
\end{equation}
or using Eddington parameter
$\Gamma_\mathrm{H}$
in a pure hydrogen plasma
(cf. Eq.\,\ref{babice})
\begin{equation}
\label{adamov}
\Gamma_\text{H}\equiv\frac{\sigma_\text{e}^\text{H} L}{4\pi c G{M}}>1,
\end{equation}
where
\begin{equation}
\sigma_\text{e}^\text{H}=\frac{n_\text{e}^\text{H}s_\text{e}}{\rho_\text{H}},
\end{equation}
$s_\text{e}$ is the Thomson scattering cross-section, and
$n_\text{e}^\text{H}$ is the electron number density in hydrogen plasma with the
density $\rho_\text{H}$.
Condition \eqref{adamov} can be rewritten in a more convenient form.
Now we
compare conditions in
pure
hydrogen wind with that in hydrogen-helium atmosphere.
Using Eqs.\,\eqref{babice} and \eqref{mezipoli} 
for the hydrogen-helium atmosphere,
%
%
we can 
express the fraction
$
%
L/\zav{4\pi c G{M}}$ in
\eqref{adamov} using the Eddington parameter in the stellar atmosphere
$\Gamma$
and we obtain
\begin{equation}
\Gamma = \Gamma_\mathrm{H}
\frac{\zav{\sigma_\mathrm{e}}_\mathrm{atmosphere}}
{\zav{\sigma_\mathrm{e}^\mathrm{H}}_\mathrm{wind}},
\end{equation}
which yields the condition ($\rho=\rho_\text{H}+\rho_\text{He}$ is the
total density 
in
the atmosphere)
\begin{equation}
\Gamma>\zav{\frac{\rho_\text{H}}{n_\text{e}^\text{H}}}
_\text{wind}
       \zav{\frac{n_\text{e}}{\rho_\text{H}+\rho_\text{He}}}
_\text{atmosphere}
.
\end{equation}
For a chemical composition given by primordial nucleosynthesis and
assuming a fully ionized gas
both in the atmosphere and in the wind
we obtain simple condition
\begin{equation}
\label{vodikpod}
\Gamma\gtrsim0.859. 
\end{equation}
Clearly, pure hydrogen-helium stars close to the Eddington limit
(with $0.859 \lesssim \Gamma < 1$)
may have
a pure
hydrogen wind,
but not the \element{H}-\element{He} wind.

The only stars from our sample for which $\Gamma\gtrsim0.859$ and which
thus can have a pure hydrogen wind are the most massive stars with
$M=300\,\Msun$ (for all considered values of effective temperature) and
star with $M=250\,\Msun$ and $ \Teff =40\,000\,\text{K}$.

However, despite this behaviour of the radiative force in pure hydrogen
models, pure hydrogen winds 
need
not necesserily exist in all stars for which
they are
%
theoretically
possible.
Radiative force in these models is not due to the hydrogen line
transitions, but mostly due to the free electrons.
Thus, in order to hydrogen wind exists, there should
be some proccess which is able to separate hydrogen and helium atoms.
The mass-loss rate of such a theoretical wind depends mainly
on the rate of this separating process. The only possible
process, which comes into consideration, is the
gravitational settling due to the different atomic masses
(see Michaud \cite{mpoprad} for a recent review).
However, the typical rate of such a process is very small, thus
mass-loss rates of a pure hydrogen wind would be probably also very
small
(see also Appendix~\ref{usazovani}).
Moreover, any macroscopic motion in the stellar atmosphere may inhibit
this gravitational settling.
Thus, we conclude that
\emph{
any possible
pure
hydrogen wind 
has either
a
very low mass-loss rate or is
even missing.}




\section{Other isotopes and elements}

\subsection{Contribution to the radiative force}

Hydrogen and helium isotopes \element[][2][1]{H} and
\element[][3][2]{He} were produced in a nonnegligible amount during the
era of the Big-Bang nucleosynthesis. 
The positions of their lines are slightly shifted with respect to those
of \element[][1][1]{H} and \element[][4][2]{He} and thus they can
potentially be exposed to a slightly higher flux than their much more
abundant counterparts. 

To test their importance we included \element[][2][1]{H} and
\element[][3][2]{He} into our atomic list.
We used basically the same atomic data for the calculation of model
atoms, however with modified energy levels due to the isotopic shift.
The isotopic abundances were taken from the simulation of primordial
nucleosynthesis by Coc et al. (\cite{coc}),
$N(
\element[][1][1]{H}
)/N(\text{\element[][2][1]{H}})=2.6 \times 10^{-5}$,
$N(
\element[][1][1]{H}
)/N(\text{\element[][3][2]{He}})=1.04 \times 10^{-5}$,
which are in a relatively good agreement with the observed values. 

The importance of other hydrogen and helium isotopes was tested using
optically thin models discussed in Sect.\,\ref{tenkovit}.
The calculated excitation and ionization states of \element[][2][1]{H}
and \element[][3][2]{He} are very similar to those of
\element[][1][1]{H} and \element[][4][2]{He}.
Moreover, due to the relatively small isotopic shifts and due to the
relatively broad photospheric lines the flux at the positions of lines
of less abundant isotopes is similar to that of more abundant isotopes.
Finally, the abundance of these isotopes is
several orders of magnitude lower than the ordinary ones, which
significantly diminishes the optically thin radiative force.
Consequently, our models showed that the contribution of less abundant
hydrogen and helium isotopes \element[][2][1]{H} and
\element[][3][2]{He} to the radiative force is negligible.

Similarly, we have also tested the contribution of lithium to the total
radiative force.
To this end,
we have selected data on excitation and ionization of lithium ions from
the Opacity Project database (Seaton \cite{top}, Peach et al.
\cite{topli}, Fernley et al. \cite{toplidva}).
The primordial lithium abundance
$N(
\element[][1][1]{H}
)/N(
\element[][7][3]{Li}
)=4.15 \times 10^{-10}$
was taken from Coc et al. (\cite{coc}).
Numerical calculation showed that due to the extremely low lithium
abundance its contribution to the radiative force can be also neglected.

\subsection{Expulsion of particular isotopes and abundance
stratification}

Slightly different radiative force acting on different isotopes and
different isotopic mass may cause radiative levitation of some isotopes
and/or gravitation settling of others, the effect also referred to as
the light induced drift (Aret \& Sapar \cite{lid}).
Thus, atmospheres of first generation stars may be chemically
stratified.
However, the discussion of these issues is beyond the scope of the
present paper which is mainly intended to the study of an outflow from
first generation stars.
There appears a more interesting question in this context.
Although the radiation force due to the less abundant isotopes is very
small and has only marginal influence on the stellar atmospheres, it might
be sufficient to expel these elements out from the stellar surface
into the interstellar medium.

However, numerical calculations showed that this is probably not the case.
Similar considerations already discussed in Sect.\,\ref{kapunikprvku}
for ordinary hydrogen and helium are valid also for deuterium and the
\element[][3][2]{He} isotope.
Radiative force acting on the deuterium is too low to allow a pure
deuterium wind.
Moreover, the radiative force due to free electrons is lower in
\element[][3][2]{He} wind due to a lower number of free electrons per
one nucleon compared to the \element{H}-\element{He} wind.

Similarly, numerical calculations
showed that pure lithium wind is also unlikely. Although the radiative force
acting on on lithium may be higher than the gravity force in deeper layers of
some models, in the outer regions is always too low to drive the lithium
wind.

\section{Influence of line transitions on accretion physics}

Up to now, the initial mass function of first stars is not known (see
Bromm \& Larson \cite{brola}).
One of the problems with the determination of initial mass function of
first stars is connected with the process of termination of accretion
onto a protostar. Dust grains are not present there, so
the radiation pressure on dust grains, which operates in
contemporary stars is not apparently available there. Thus,
other mechanisms may be important for metal-free stars (e.g. Omukai \&
Inutsuka \cite{omin}, Tan \& McKee \cite{tancee}).

Here we do not intend to study accretion physics in detail, but only to
depict such processes which may be important for termination of
accretion, namely we would like to test whether line transitions due to
hydrogen and helium are able to influence the accretion.

Thus, we selected a fixed velocity law which can be obtained from the
solution of the momentum equation for isothermal ($T=T_\mathrm{wind}$)
spherical accretion (Mihalas \cite{mihalas})
\begin{equation}
\label{padani}
\zav{\vr-\frac{a^2}{\vr}}\frac{\de \vr}{\de r}=
\frac{2a^2}{r}-g.
\end{equation}
The boundary condition of Eq.~\eqref{padani} was artificially selected
to be $\vr(100R_*)=\sqrt{\frac{5}{3}}a$ and the density structure was
calculated from the continuity equation
\begin{equation}
\label{konti}
4\pi\rho\vr r^2=\dot M=\text{const.},
\end{equation}
where $\dot M$ is the accretion rate.
We selected three different values of accretion rates, namely
$10^{-3}M_\odot\,\text{year}^{-1}$, $10^{-5}M_\odot\,\text{year}^{-1}$,
$10^{-7}M_\odot\,\text{year}^{-1}$.
The highest value of accretion rate roughly corresponds to the critical
accretion rate over which protostellar core reaches the Eddington limit
and no supermassive star forms (Omukai \& Palla \cite{ompal}).

For the velocity and density structure mentioned above we solved
statistical equilibrium equations for hydrogen and helium inflow.
We neglected any influence of possible shock at the stellar surface and
assumed that the protostar is in hydrostatic equilibrium.
In such a case for our numerical calculations
we may use the same stellar parameters as we used for wind tests, namely
the stellar parameters taken from Kudritzki (\cite{kudmet}). 
Our numerical calculations showed that the sum of the
radiative force in the Sobolev
approximation calculated after Eq.\,\eqref{zarsil} and
the radiative force acting on free electrons can together exceed the
gravitational force only in stars close to the Eddington limit and only
for relatively small accretion rates (of the order
$10^{-5} \Msun \,\text{year}^{-1}$ and smaller).

The principal difference between possible stellar wind and accretion
flow is that important regions of accretion occur at much larger
distances from the star than the acceleration region of the 
\zm{stellar}
wind.
Since the ionizing stellar flux at larger distances is lower, the
inflow (especially helium) is less ionized and the line radiative force
due to \ion{He}{ii} becomes important
%
relative to the gravity
force.

Clearly, the importance of the radiative force rises for more massive
stars closer to the Eddington limit.
Thus, there is a possibility that for even more massive stars than those
considered here the radiative force both due to line transitions and
free electrons is able to overcome gravity and set the initial mass
function of supermassive stars.
These considerations have to be, however, tested against more advanced
accretion models.

\section{Discussion of simplifying assumptions}
%

\subsection{Radiative transfer}
There are several problems which can influence our final results.
Probably the most serious one is the simplified radiative transfer in lines.
This can influence the calculated occupation numbers and hence also the
radiative force.
However, this is probably not the case, since the results obtained are
relatively robust, i.e. in many cases even the optically thin radiative
force is not sufficient to drive a stellar wind.

\subsection{Level dissolution}
There is another more subtle problem which is related to the neglect
of the so-called level dissolution (Hummer \& Mihalas \cite{ocupro}). 
It is the statistical treatment of the fact that the highest atomic
levels cease to exist under the influence of surrounding particles.
Usually this problem is treated approximately using the lowering of
ionization potential.
However, it
can be treated in a consistent way using the so-called
occupation probability formalism (Hubeny et al. \cite{huhul},
Kub\'at \cite{ATA3}).
Possibly, neglect of level dissolution can influence the calculated
radiative force, specially in the case of hydrogen.
However, test calculations with different number of hydrogen and helium
lines showed this is not the case.
Although total number densities of individual ionization states may
differ, number densities of lower excited levels (which are important
for line driving) are nearly the same.
Since there is a relatively low radiative flux in the wavelength regions
of lines originating from higher excited levels, their contribution to
the radiative force is negligible (see also Sect.\,\ref{vodikprik}).
Thus, we conclude that the influence of level dissolution on our results
is only marginal.


\subsection{Pulsations}
%

Since we assume stationary flow here, the effect of possible stellar
pulsations is neglected.
According to numerical simulations of Baraffe et al. (\cite{bahewo}),
primordial stars may actually lose mass via pulsational instability.
However, such effect is connected mostly with the stellar interior, so
it is beyond the scope of 
\zm{the present}
paper.


\section{Conclusions}

\subsection{Stars with chemical composition given by primordial nucleosynthesis}

We have studied the dynamical aspects of a possible circumstellar
envelope around massive first stars using our NLTE models.
Although many of studied stars are very close to the Eddington limit and the
maximum possible radiative force (calculated assuming that all lines are
optically thin) is high in some cases,
the stellar winds driven by line transitions are unlikely, because the
line radiative force is very small.
The influence of other isotopes and elements, which had been
produced during primordial nucleosynthesis, was found to be marginal.
Also the effect of the continuum radiative force is negligible.

We have also tested whether first stars can expel individual elements,
either the dominant ones (i.e. hydrogen and helium) or the trace ones
(i.e. hydrogen and helium isotopes or lithium).
Calculations showed that the only theoretically possible chemically
homogeneous outflow is that of hydrogen
(for stars very close to the Eddington limit with $\Gamma\gtrsim0.859$),
%
since other elements
yield
a small number of free electrons per nucleon (and the line-radiative force is
always negligible). However, even the case of a pure
hydrogen outflow is unlikely since the
radiative force in such a wind would be predominantly due to the free
electrons. Hence, another process which would be able to separate hydrogen
and helium atoms in the atmosphere should
exists (e.g.~gravitational settling) to enable a pure hydrogen wind.
Without such a process a pure hydrogen wind is not possible.

%

Finally, we have tested whether the radiative force due to line
transitions is able to terminate the initial accretion onto first stars.
We have found that this is only the case of stars relatively close to
the Eddington limit and with relatively small accretion rates.
Consequently, it is possible that also
for even more massive stars this process
can influence the accretion physics
and
%
the initial mass-function of first stars.
%

\zm{%
Our results do not imply that the first stars
are completely without line driven winds
during all of their life time.}
For example, cooler stars (which were not inlcuded into our present
study) could have wind due to the hydrogen or helium lines. Moreover,
if the accretion terminates, then as soon as sufficient amount of
heavier elements is synthesised in the stellar core and transported to
the surface layers, the line-driven wind is initiated
(Kudritzki \cite{kudmet}).
If this happens at a sufficiently low metallicity, such a stellar wind
is multicomponent, allowing for element separation and frictional
heating (Krti\v cka et al. \cite{gla}).

\subsection{General picture of stellar wind of massive low-metallicity
stars}

From this study and previous ones (Kudritzki
\cite{kudmet}, Krti\v{c}ka et al. \cite{gla})
a general picture of the properties of stellar
wind of massive stars at low metallicities emerges.
\begin{enumerate}
\renewcommand{\labelenumi}{\roman{enumi}}
\item
\label{bezvi}
Zero-metallicity stars with $\Gamma\lesssim0.859$ likely do not have any
stellar wind.
If their atmospheres are relatively quiet (i.e. in the absence of
violent pulsations or fast rotation) the chemical
peculiarity in their atmospheres may develop.
On the other hand, zero-metallicity \element{H}-\element{He}
stars with $\Gamma\gtrsim0.859$ may have
pure-hydrogen wind
(probably very thin
with mass-loss rates of order 
$10^{-14}\Msun\,\text{year}^{-1}$, see Appendix~\ref{vodikslepak}),
%
driven only by the light scattering on free electrons.
%
\item
\label{kovovi}
With increasing metallicity the line-driven wind may develop.
For extremely low metallicities such a stellar wind is likely pure
metallic (see Babel \cite{babelb}).
\item
For higher metallicities also hydrogen and helium are expelled from the
atmosphere and multicomponent effects become important in such a stellar
wind (Krti\v{c}ka et al. \cite{gla}).
The multicomponent structure does not significantly influence the wind
mass-loss rate, which may be approximated by formulas obtained for
one-component flow by Kudritzki (\cite{kudmet}).
\item
For higher metallicities, a
line-driven wind which is similar to the stellar wind of present hot
stars exists and its mass-loss rate was calculated by Kudritzki
(\cite{kudmet}). 
\end{enumerate}
Thus, in our opinion, the best choice for the evolutionary calculation
now is to assume zero mass-loss rate for
zero and low metallicity stars
mentioned in items
i--ii
and to use Kudritzki (\cite{kudmet})
prescription for other stars.

\begin{acknowledgements}
We would like to thank to Dr. S. Owocki for pointing our attention
to this problem.
This research has made use of NASA's Astrophysics Data System.
This work was supported by grants GA \v{C}R 
205/02/0445, 205/03/D020, 205/04/1267.
The Astronomical Institute Ond\v{r}ejov is supported by a project Z1003909.
\end{acknowledgements}

\appendix
\section{Gravitational settling time scale}\label{usazovani}
%
Let us at least roughly
estimate
the characteristic time-scale
$\tau_\text{D}$ for gravitational settling of helium in the atmospheres
of first stars.
This is given by (Kippenhahn \&
Weigert
\cite{kiwe}, p.\,59)
\begin{equation}
\tau_\text{D}\approx\frac{S^2}{D},
\end{equation}
where $S$ is the characteristic length for the density variation and the
diffusion coefficient
\begin{equation}
D\approx\ell\sqrt{\frac{kT}{3\vo m}}.
\end{equation}
The mean free path
of helium in hydrogen plasma
%
$\ell$ we 
estimate
using
the
expression for Coulomb collision frequency (Burgers \cite{burgers}).
For typical parameters of the outer parts of the atmosphere
($S=10^{11}\,\text{cm}$, $n\approx10^8\,\text{cm}^{-3}$,
$T\approx30\,000\,\text{K}$) we obtain mean-free path of
$\ell\approx10^4\,\text{cm}$ and corresponding characteristic diffusion
time scale $\tau_\text{D}\approx10^{4}\,\text{years}$, which shows that
the gravitational settling is really possible in the outer parts of
atmospheres of first stars.
On the
other hand, in deeper atmospheric layers (where $S=10^{11}\,\text{cm}$,
$n\approx10^{13}\,\text{cm}^{-3}$, $T\approx50\,000\,\text{K}$) we
obtain mean-free path of $\ell\approx1\,\text{cm}$ and the diffusion
time scale $\tau_\text{D}\approx10^{8}\,\text{years}$, which highly
exceeds the expected lifetime of massive first stars (Marigo et al.
\cite{mari}), which is of order $10^{6}\,\text{years}$.
Clearly, in the deeper layers the gravitational settling of helium
probably does not take place.

Anyway, the previous analysis enables us to roughly estimate the
mass-loss rate due to the pure hydrogen wind, which is
\begin{equation}
\vo{\dot M}\approx4\pi R_*^2Sn\vo m\tau_\text{D}^{-1}.
\end{equation}
Note that since $D\sim 1/n$ the estimated mass-loss rate does not depend
on the atmospheric number density $n$, thus for both atmospheric points
mentioned above we obtain
$\vo{\dot M}\approx\,10^{-16} \text{M}_\odot\,\text{year}^{-1}$.
This value is close to a more exactly estimated upper limit of such
hydrogen stellar wind, which is derived in Appendix~\ref{vodikslepak}.

\section{Pure hydrogen wind driven by free electrons}
\label{vodikslepak}

The radiative acceleration due to the light scattering on free electrons
may be higher than the gravitational acceleration in the pure hydrogen
plasma for stars with $\Gamma\gtrsim0.859$ (see Eq.\,\eqref{vodikpod}).
Thus, if the hydrogen layer builds up at the surface of such star (e.g.
due to the gravitational settling) then pure hydrogen wind may occur
(cf. Malov \cite{nemam} for a description of stellar wind driven by
Thomson scattering).
In a stationary state such a stellar wind may be described (assuming
spherical symmetry) by hydrodynamical equations (Burgers \cite{burgers},
see also Krti\v{c}ka \& Kub\'at \cite{kkii})
\begin{subequations}
\label{samvodikvitr}
\begin{gather}
\label{samvoko}
\frac{\de}{\de r}\zav{r^2\vo\rho{\vo\vr}}  =  0,\\
\label{samvopo}
{\vo\vr}\frac{\de {\vo\vr}}{\de r}= g^\text{rad}_\text{e,\,H}-g-
   \frac{1}{\vo\rho}\frac{\de}{\de r}\zav{\vo a^2\vo\rho}
   -\frac{1}{\vo\rho} K_{\text{H}\alpha}G(x_{\text{H}\alpha}),
\end{gather}
\end{subequations}
where indexes $\mathrm{H}$ and $\alpha$ stand for hydrogen and alpha
particles, $\vo\vr$ and $\vo\rho$ are the radial velocity and the density of 
completely ionized pure hydrogen
plasma, $\vo a^2=2kT/\vo m$ (we assume that hydrogen and helium
temperares are equal), the frictional parameter
\begin{equation}
K_{\text{H}\alpha}=\vo n n_\alpha k_{\text{H}\alpha}=
                   \vo n n_\alpha \frac{4\pi\vo q^2q_\alpha^2}{k T}\ln\Lambda,
\end{equation}
and where the Chandrasekhar function $G(x)$, defined in terms
of the error function $\erf(x)$, is
\begin{equation}
G(x)=\frac{1}{2 x^2}\zav{\erf(x)-\frac{2x}{\sqrt\pi}\exp\zav{-x^2}},
\end{equation}
and the argument of the Chandrasekhar function is
\begin{equation}
x_{\text{H}\alpha}=\frac{\vo\vr}{\alpha_{\text{H}\alpha}},
\end{equation}
where we have assumed that helium is static in the atmosphere and
\begin{equation}
\alpha_{\text{H}\alpha}^2=\frac{2kT\zav{\vo m+m_\alpha}} {\vo m m_\alpha}.
\end{equation}

Inserting continuity equation Eq.~\eqref{samvoko} into Eq.~\eqref{samvopo} and
using the definition of $\Gamma_\text{H}$ (see Eq.~\eqref{adamov}) we
obtain momentum equation in the form of
\begin{multline}
\label{rovod}
\zav{\vo\vr-\frac{\vo a^2}{\vo\vr}}\frac{\de {\vo\vr}}{\de r}=\\=
\frac{GM}{r^2}\zav{\Gamma_\text{H}-1}-\frac{\de \vo a^2}{\de r}+\frac{2\vo a^2}{r}-
\frac{n_\alpha}{\vo m}k_{\text{H}\alpha}G(x_{\text{H}\alpha}).
\end{multline}
This equation has a critical point for $\vo\vr=\vo a$. To obtain
a smooth solution at this critical point the critical condition
\begin{equation}
\label{krivo}
\frac{GM}{r^2}\zav{\Gamma_\text{H}-1}-\frac{\de \vo a^2}{\de r}+
\frac{2\vo a^2}{r}-
\frac{n_\alpha}{\vo m}k_{\text{H}\alpha}
G\zav{\frac{\vo a}{\alpha_{\text{H}\alpha}}}=0
\end{equation}
shall be fulfilled.
Finally, the helium density in a static case is given by the
equation of hydrostatic equilibrium in the form of
\begin{equation}
\label{rohel}
-\frac{GM}{r^2}\zav{1-\Gamma_\alpha}-
\frac{1}{\rho_\alpha}\frac{\de}{\de r}\zav{a_\alpha^2\rho_\alpha}
+\frac{\vo n}{m_\alpha}k_{\text{H}\alpha}G(x_{\text{H}\alpha})=0.
\end{equation}
Eqs.\,\eqref{rovod}--\eqref{rohel} (together with boundary condition
which links hydrogen and helium density deep in the stellar atmosphere)
can be used to calculate wind models of pure hydrogen flow.
The mass-loss rate of such wind can be obtained by the requirement that
the critical condition \eqref{krivo} is fulfilled at the critical point.

We do not aim to solve equations for pure hydrogen wind here, however we present
only an estimate of upper limit of the mass-loss rate.
To make its derivation more tractable, we neglect gas-pressure terms in
Eqs.~\eqref{rovod}--\eqref{rohel}. The critical condition \eqref{krivo} then
yields the helium density at the critical point as
\begin{equation}
n_\alpha^\text{crit}\approx \frac{GM}{r_\text{crit}^2}\zav{\Gamma_\text{H}-1} 
        \frac{\vo m}{k_{\text{H}\alpha}}
	\hzav{G\zav{\frac{\vo a}{\alpha_{\text{H}\alpha}}}}^{-1}.
\end{equation}
Similarly, the gas
pressure term in the helium hydrostatic equation Eq.\,\eqref{rohel} is
always possitive, thus the upper limit for hydrogen density at the
critical point is
\begin{equation}
\vo n^\text{crit}\lesssim\frac{GM}{r_\text{crit}^2}\zav{1-\Gamma_\alpha}
        \frac{m_\alpha}{k_{\text{H}\alpha}}
        \hzav{G\zav{\frac{\vo a}{\alpha_{\text{H}\alpha}}}}^{-1}.
\end{equation}
With this we can derive upper limit for hydrogen mass-loss rate as
\begin{multline}
\vo{\dot M}\lesssim4\pi\,\vo n^\text{crit}\,\vo\vr^\text{crit}\,r_\text{crit}^2
=\\= 4\pi\vo aGM\zav{1-\Gamma_\alpha} \frac{\vo m m_\alpha}{k_{\text{H}\alpha}}
     \hzav{G\zav{\frac{\vo a}{\alpha_{\text{H}\alpha}}}}^{-1}.
\end{multline}
Assuming $q_\alpha=2\vo q$ this can be rewritten as
\begin{equation}
\vo{\dot M}\lesssim
     2\times10^{-14}\, \text{M}_\odot\,\text{year}^{-1}\!
     \zav{\frac{M}{100\,\text{M}_\odot}}\!\!\zav{\frac{T}{10^4\,\text{K}}}^{3/2}
     \!\!\zav{1-\Gamma_\alpha}.
\end{equation}
This shows that the mass-loss rate of hypothetical pure hydrogen wind is very
small.

The wind velocity of such a wind outside the stellar atmosphere can be easily
obtained from Eq.~\eqref{rovod} assuming negligible frictional force
($G(x_{\text{H}\alpha})\ll1$), since in such a case we obtain equation for the
Parker's type stellar wind (assuming isothermal wind, see Mihalas~\cite{mihalas}). Note however,
that the wind terminal velocity is influenced either by the radiative
heating and cooling processes (which influence the sound speed $\vo a$) 
or
by
hydrogen recombination, which decreases the radiative force on free electrons.

Note however that the real situation will be likely more complex than that
presented here. First, Eqs.~\eqref{samvodikvitr} were derived assuming
maxwellian distributions of velocities for all the components. While this may be
true for hydrogen and helium, the velocity distribution of electrons may have
two maxima (one corresponding to electrons which move together with hydrogen and
one corresponding to electrons which stay in the atmosphere). Possibly, also
electric polarisation field, which was neglected in Eqs.~\eqref{samvodikvitr},
may play some role. Finally, the process described here may not be stationary.
Anyway, it is not likely that pure hydrogen mass-loss rate is by orders of
magnitude higher than that presented here, hence the influence of pure hydrogen
mass-loss on the evolution of first stars is only marginal.

For the calculation of the radiative force we have assumed that the stellar wind is
optically thin for the Thomson scattering, which seems to be a plausible
approximation given the small mass-loss rate.
%

It may seem surprising that we propose here that wind driven by
scattering on free electrons may
exist,
whereas Malov (\cite{nemam}) concluded that such wind may exist only in
the case of
a
positive temperature gradient $\de T/\de r >0$ at the wind critical
point.
His conclusion is based on critical point condition \eqref{krivo}, which
indeed cannot be fulfiled if $\de T/\de r \leq0$ and if we do not
account
for
friction due to the helium atoms.
Here, however, the friction due to the helium atoms enables to fulfill
the critical point condition and acts in a similar way as
positive
temperature gradient in
Malov's
(\cite{nemam}) paper.

\newcommand{\MSA}[1]{in Modelling of Stellar Atmospheres, IAU Symp. 210,
        N. E. Piskunov, W. W. Weiss \& D. F. Gray eds., Astron. Soc.
        Pacific, #1}

\end{document}